\RequirePackage{rotating}
\documentclass[10pt]{iopart}

\usepackage{pbox}
\usepackage{graphicx}
\usepackage{dcolumn}
\usepackage{bm}
\usepackage[square,sort&compress,numbers]{natbib}
\usepackage{hyperref,outlines,enumitem,rotating}
\usepackage{tikz}
\usetikzlibrary{decorations.pathmorphing}
\expandafter\let\csname equation*\endcsname\relax
\expandafter\let\csname endequation*\endcsname\relax
\usepackage{amsmath,amssymb}
\usepackage{setspace}
\hypersetup{colorlinks=true,urlcolor=blue}
\usepackage{afterpage}

\begin{document}

\title{Classical Tools for Antipodal Identification in Reissner-Nordstr\"om Spacetime}

\author{Nathaniel A. Strauss$^1$, Bernard F. Whiting$^1$ and Anne T. Franzen$^2$}
\address{$^1$ Department of Physics, University of Florida, Gainesville, Florida, USA}
\address{$^2$ Center for Mathematical Analysis, Geometry and Dynamical Systems, Mathematics Department, Instituto Superior T\'ecnico, Universidade de Lisboa, Lisbon, Portugal} 
\eads{\mailto{straussn@ufl.edu}, \mailto{bernard@phys.ufl.edu}, \mailto{anne.franzen@tecnico.ulisboa.pt}}

\date{\today}

\begin{abstract}
We extend the discussion of the antipodal identification of black holes to the Reissner-Nordstr\"om (RN) spacetime by developing the classical tools necessary to define the corresponding quantum field theory (QFT). We solve the massless Klein-Gordon equation in the RN background in terms of scattering coefficients and provide a procedure for constructing a solution for an arbitrary analytic extension of RN. The behavior of the maximally extended solution is highly dependent upon the coefficients of scattering between the inner and outer horizons, so we present the low-frequency behavior of, and numerical solutions for, these quantities. We find that, for low enough frequency, field amplitudes of solutions with purely positive or negative frequency at each horizon will acquire only a phase after passing both the inner and outer horizons, while at higher frequencies the amplitudes will tend to grow exponentially either to the future or to the past, and decay exponentially in the other direction. Regardless, we can always construct a basis of globally antipodal symmetric and antisymmetric solutions for any finite analytic extension of RN. We have characterized this basis in terms of positive and negative frequency solutions for future use in constructing the corresponding QFT.
\end{abstract}
\noindent{\it Keywords\/}: Reissner-Nordstr\"om, scalar field, quantum field theory, general relativity, black hole

\maketitle

\section{Introduction}

Antipodal identification is the physical identification of, or a mapping between, two regions of spacetime on a manifold, where the precise meaning of ``antipodal'' varies by the manifold. Interest in antipodal identification began in the construction of quantum field theory (QFT) in four-dimensional de Sitter space, proposed by Schr\"odinger in 1957 and explicitly developed by Sanchez in 1987~\citep{schroedinger2011,sanchez1987_2}. In this so called ``elliptic interpretation'' two spacelike separated points are physically identified as a single space point with the future light cone of one of the points coinciding with the past light cone of the other point. As has subsequently been noticed by Gibbons and others, the ``elliptic interpretation'', which may arise in connection with discussions on topology change \cite{Gibbons:1991ey,Barvinsky:2012qm}, could have an impact on quantum mechanics \cite{Gibbons:1986dd}, can lead to lack of time orientation  \cite{Gibbons:1993iv} and the absence of a spinor structure \cite{Gibbons:1991tp}, and can result in causality violation \cite{Gibbons:1999uv}.
Nevertheless, Sanchez proposed a QFT for this model in terms of antipodal symmetric fields, though not requiring the states to have symmetry. This is finally compared to scenarios without identified points. She had used these ideas earlier in a collaboration with Castagnino et al., in which they investigated constant curvature, self-consistent solutions of semiclassical Einstein equations~\citep{castagnino1987}. Further, together in Domenech et al., they extend this analysis for anti-de Sitter and Rindler spacetimes~\citep{domenech1987}. Later, Stephens supported the antipodal identification in Rindler spacetimes by demonstrating its importance in the interpretation of pair production in a uniform electric field~\citep{stephens1989}. Friedman and Higuchi then explained how time non-orientable QFT can be constructed locally in antipodally identified de Sitter spacetime by viewing the spacetime as a four-dimensional M\"obius strip~\citep{friedman1995}. In contrast to~\citep{sanchez1987_2}, they employ antipodally symmetric algebraic states on de Sitter space. Alternatively, Chang and Li found that de Sitter spacetime with antipodal identification is in fact orientable in odd dimensions, and they explicitly construct the resulting QFT~\citep{chang2007}.

The discussion of antipodal identification becomes much richer in black hole spacetimes, and most of the discussion so far has been confined to the Schwarzschild (charge-neutral, non-rotating) black hole. In concurrence with work on de Sitter space, motivated by unconventional ideas of 't Hooft~\citep{tHooft:1983kru,tHooft:1984kcu}, Sanchez and Whiting~\citep{sanchez1987} explicitly constructed a globally antipodally symmetric scalar field on the Schwarzschild manifold. They found that the usual Fock space construction is not possible and the usual thermal features are not present (although Martellini and Sanchez have proposed a construction which circumvents this difficulty \cite{Martellini:1987ug}). Subsequently, Whiting discussed the effect of these explicit representations of symmetric fields on the role of gravitation in thermal physics ~\citep{whiting1989}. More recently, 't Hooft has suggested an antipodal identification of the Schwarzschild black hole motivated by the quantum information  paradox~\citep{'thooft2016,'thooft2016_2,'thooft2017,'thooft2018}. The resulting unusual structure of spacetime restores quantum pureness and time reversal symmetry,  properties usually not seen in thermodynamic objects. In~\citep{'thooft2016} he first proposes that unitarity should be assumed as a crucial guiding principle, which leads to a topology of Schwarzschild spacetime different from the one usually assumed.  According to 't Hooft, Hawking radiation is only locally thermal. Globally however, quantum states evolve purely unitarily, assuming an entanglement of each particle in the first exterior with a partner appearing in the second exterior, and similarly for particles in the interior. He refers to this entanglement as antipodal identification. 't Hooft also explains how this causes a topological twist of the background metric and even removes firewalls~\citep{'thooft2016_2}.  He explains this in more detail in~\citep{'thooft2017}, while a shorter review of the idea is given in~\citep{'thooft2018}. The effects of this entanglement on black hole scattering and vacuum states has been investigated~\citep{betzios2016,bzowski2018}. Antipodal identification has also found recent support in the construction of theories of quantum gravity~\citep{sanchez2018,sanchez2019}. In sum, it has been shown that antipodal identification has a large effect on the quantum and thermal features of black holes.

While quantum field theories on the domain of outer communication of black holes with charge and/or angular momentum have been extensively studied (e.g.~\citep{crispino2009}), field theories on the the analytic extensions necessary for antipodal identification are less well studied, particularly for the maximal analytic extension. The contribution of this paper is to extend the discussion of antipodal identification to the Reissner-Nordstr\"om (RN) spacetime, that of a charged, non-rotating black hole. Proceeding in the same manner as Sanchez and Whiting~\citep{sanchez1987}, we solve the massless Klein-Gordon equation on the RN background, and we construct antipodal symmetric fields in terms of positive- and negative-frequency solutions, which could be used in the antipodal identification of the corresponding QFT (as done by others, e.g.~\citep{schroedinger2011,sanchez1987_2}). Two primary difficulties arise. First, the presence of more than one horizon requires the consideration of more than one antipodal identification on the analytic extension of the RN geometry. Constructing symmetric fields about a single horizon is analogous to the Schwarzschild case. Constructing a solution simultaneously symmetric about multiple horizons is also possible if we utilize solutions with positive frequency at one horizon and negative frequency at another horizon. Second, the amplitude of a solution in an arbitrary analytic extension is highly dependent upon the scattering coefficients for propagation between the two horizons of RN. For this reason we examine the low-frequency limit of these coefficients, as well as their values obtained from a numerical series approximation. We find that for certain frequency ranges of the scalar field, and especially at high frequency, the amplitude of the field grows as it propagates through the maximally extended geometry. Despite this fact, construction of the antipodal symmetric fields, as might be used to define the QFT, is possible for an analytic extension of any finite number of horizons.

The structure of the paper is as follows. In section~\ref{sec:RN_manifoild}, we introduce the RN manifold and its Kruskal-Szekeres coordinates. Further, we specify its characteristic regions separated by outer and inner horizons and show their representation in Penrose diagrams. In section~\ref{sec:klein_gordon} we solve the massless Klein-Gordon equation in each spacetime region and demonstrate how to build global solutions. In section~\ref{sec:scattering}, we present the low frequency limits of, and numerical solutions for, the coefficients of scattering between the two RN horizons. In section~\ref{sec:linear_independence_symmetry} we discuss the linear independence of Klein-Gordon solutions in an analytical extension of RN and demonstrate how to construct antipodal symmetric fields from these solutions. Finally, in section~\ref{sec:discussion} we summarize and discuss our results.

\section{The Reissner-Nordstr\"om Manifold}
\label{sec:RN_manifoild}

The physics of any spacetime manifold is determined by the metric $g_{\mu\nu}(x)$ and its derivatives, and the Einstein equations relate $g_{\mu\nu}$ to the distribution of matter:
\begin{align}
\label{eq:EF}
R_{\mu\nu}-\tfrac{1}{2}Rg_{\mu\nu}=8\pi T_{\mu\nu},
\end{align}
where $R_{\mu\nu}$ is the Ricci tensor, $R$ is the curvature scalar, $g_{\mu\nu}$ is the metric, and $T_{\mu\nu}$ is the stress-energy tensor. The Schwarzschild metric is a spherically symmetric non-trivial vacuum ($T_{\mu\nu}=0$) solution to~\eqref{eq:EF}, representing a charge-neutral, non-rotating black hole. The RN manifold is a generalization of the Schwarzschild manifold that allows for a spherically symmetric electromagnetic field. The assumption of spherical symmetry is retained from the Schwarzschild derivation, but on the right hand side one uses the electromagnetic stress energy tensor:
\begin{align}
T_{\mu\nu}=F_{\mu\rho}F_{\nu}{}^{\rho}-\tfrac{1}{4}g_{\mu\nu}F_{\rho\sigma}F^{\rho\sigma},
\end{align}
where $F_{\mu\nu}$ is the electromagnetic field tensor ~\citep{poisson2004}. If we assume no magnetic monopoles are present, we can choose a gauge such that the only non-vanishing component of the electromagnetic field tensor is $F^{tr}=-F^{rt}=Q/r^2$ in spherical coordinates~\citep{poisson2004}. Solving the resulting Einstein equations yields the RN manifold, which describes a charged, non-rotating black hole.

In the following sections, we examine the properties of the RN manifold in both standard spherical coordinates and in Kruskal coordinates, each of which provide valuable physical intuition for the structure of the RN manifold and the behavior of nearby particles.

\subsection{Spherical Coordinates}

The RN metric outlined above in standard spherical coordinates is
\begin{align}
g_{\mu\nu}\rmd x^\mu \rmd x^\nu = -f(r)\rmd t^2+f(r)^{-1}\rmd r^2+ r^2\rmd \theta^2+r^2\sin^2\theta\, \rmd \phi^2,
\end{align}
where
\begin{align}
f(r)=1-\frac{2M}{r}+\frac{Q^2}{r^2}=\frac{(r-r_+)(r-r_-)}{r^2},
\label{eq:orig_f}
\end{align}
where
\begin{align}
r_\pm=M\pm \sqrt{M^2-Q^2}.
\end{align}
We interpret $M$ as the mass and $Q$ as the charge of the black hole. We recover the Schwarzschild metric at $Q=0$.

We see that the $g_{rr}$ component of the metric is singular at $r=0, r_\pm$. The singularity at $r=0$ is a geometric singularity as in the Schwarzschild case, $r=r_+$ is the outer event horizon (corresponding to the Schwarzschild horizon since $r_+\rightarrow 2M$ as $Q\rightarrow 0$), and $r=r_-$ is a Cauchy horizon (which has no Schwarzschild analog, since $r_-\rightarrow0$ as $Q\rightarrow0$ and vanishes completely when $Q=0$).

Qualitatively speaking, classical geodesics in this spacetime have different behavior from those of a Schwarzschild black hole. After crossing $r_+$ from the outside, uncharged particles inevitably move inward, as in the Schwarzschild case. However, after crossing $r_-$, the singularity repels the particle and the particle's geodesic again crosses $r_-$, after which the particle inevitably exits $r_+$ again, ejected into a new external world ~\cite{poisson2004}.

\subsection{Kruskal-Szekeres Coordinates}
In the following we will introduce two sets of Kruskal-Szekeres coordinates, each of which is regular at one of the horizons. First, we define the so-called tortoise coordinate $r_*$:
\begin{align}
\frac{\rmd r_*}{\rmd r}=\frac{1}{f(r)},
\label{eq:tortoise}
\end{align}
and when integrated
\begin{equation}
r_*=\int_0^r \frac{\rmd r'}{(1-\tfrac{r_+}{r'})(1-\tfrac{r_-}{r'})} =r+\frac{1}{2\kappa_+}\ln\Big\lvert \frac{r}{r_+}-1\Big\rvert-\frac{1}{2\kappa_-}\ln\Big\lvert\frac{r}{r_-}-1\Big\rvert,
\end{equation}
where
\begin{align}
\kappa_\pm=\frac{1}{2}\Big(\frac{r_+-r_-}{r_\pm^2}\Big)
\end{align}
are the surface gravities at the outer and inner horizons, defined here to be positive. Note the following corresponding limits of $r$ and $r_*$:
\begin{align}
\begin{tabular}{c|c|c|c|c|c}
$r\rightarrow$ & $-\infty$ & $0$ & $r_-$ & $r_+$ & $\infty$\\
\hline
$r_*\rightarrow$ & $-\infty$ & $0$ & $\infty$ & $-\infty$ & $\infty$\\
\end{tabular}.
\label{eq:rstar_limits}
\end{align}

\begin{figure}
\centering
\begin{minipage}{.49\textwidth}
\centering
\includegraphics[scale=.77]{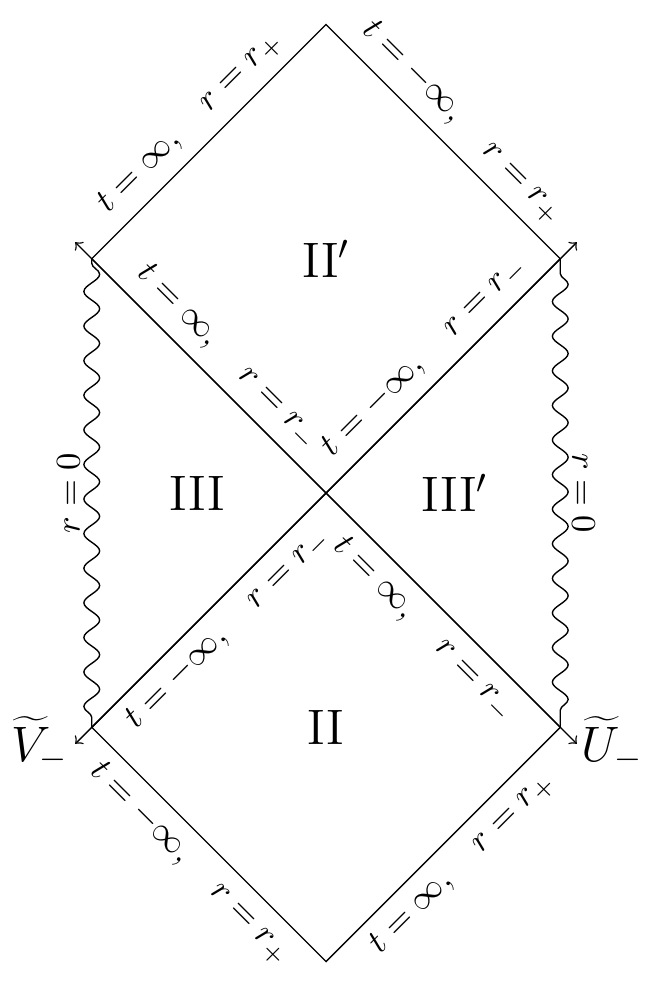}
\end{minipage}
\begin{minipage}{.49\textwidth}
\centering
\includegraphics[scale=.77]{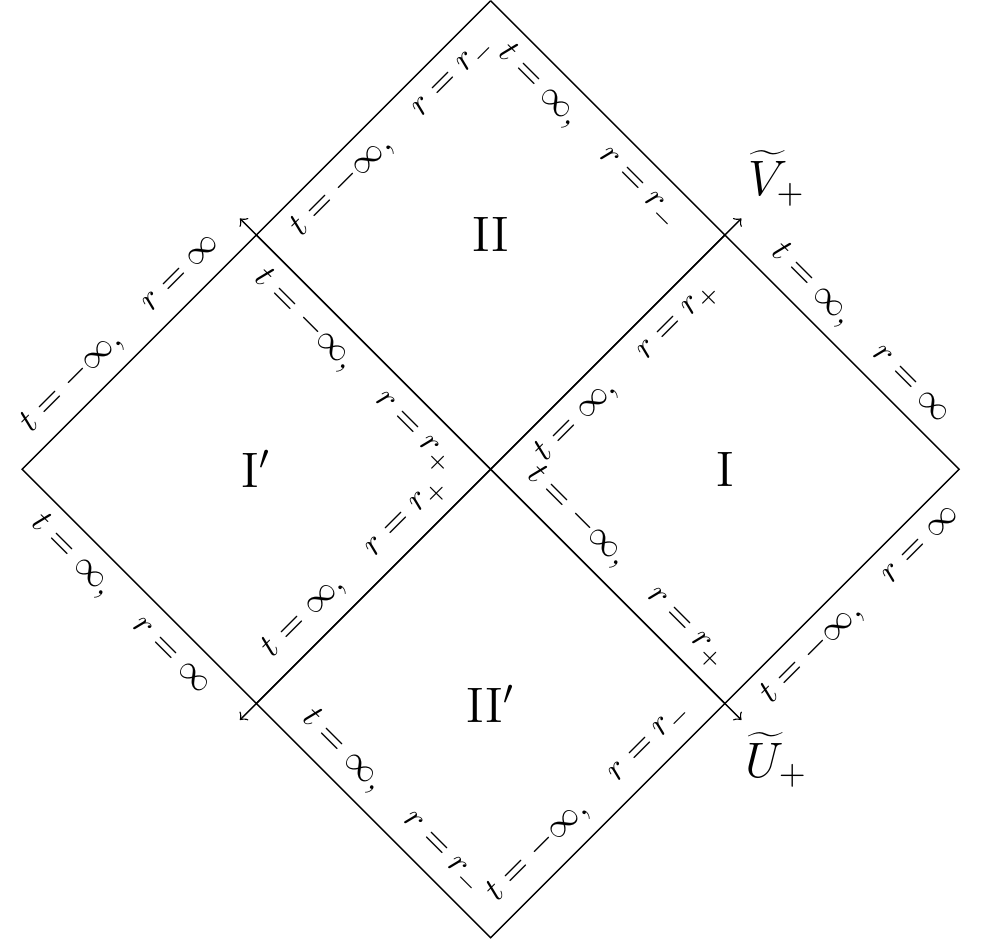}
\end{minipage}
\caption{These are the Penrose diagrams for the RN spacetime regions in the $\widetilde{U}_-\widetilde{V}_-$-plane (left) and $\widetilde{U}_+\widetilde{V}_+$-plane (right). The $r=0$ curve in the $\widetilde{U}_-\widetilde{V}_-$-plane is represented as a squiggly vertical line, though its true shape is some other curve in the $\widetilde{U}_-\widetilde{V}_-$-plane. }
\label{fig:inner_outer_coordinates}
\end{figure}

There are two sets of Kruskal-Szekeres coordinates for the RN geometry, one regular at each horizon. We keep the coordinates $\theta$ and $\phi$, while transforming $t$ and $r$ into new coordinates $U$ and $V$. The (partial) set of coordinates regular at the outer horizon is
\begin{align*}
U_+=\begin{cases}
\rme^{\kappa_+ (r_*-t)}& r>r_+\\
-\rme^{\kappa_+ (r_*-t)}& r_-<r<r_+
\end{cases}, \qquad
V_+= \rme^{\kappa_+(r_*+t)}.
\end{align*}
The change in sign of $U_+$ at $r_+$ keeps the coordinate transformation one-to-one. We call the spacetime region where $r>r_+$ region I and the spacetime region where $r_-<r<r_+$ region II. We can also introduce a sign change in the $V_+$ coordinate, where the negative values of $V$ correspond to another universe with its own $r$ and $t$ coordinates. We have then created two new spacetime regions I$'$ and II$'$, as seen in figure~\ref{fig:inner_outer_coordinates}. In order to line up the corresponding boundaries of the primed and unprimed universes, we also need to introduce a sign change in $U_+$ for the primed universe. The full set of Kruskal-Szekeres coordinates centered at the outer horizon is then
\begin{align}
U_+&=\begin{cases}
\rme^{\kappa_+ (r_*-t)}& \text{I},\text{ II}'\\
-\rme^{\kappa_+ (r_*-t)}& \text{I}',\text{ II}
\end{cases}, \label{eq:outer_coordinates_u}\\
 V_+&= \begin{cases}
\rme^{\kappa_+(r_*+t)} & \text{I, II}\\
-\rme^{\kappa_+(r_*+t)} & \text{I}',\text{ II}'
\end{cases}.
\label{eq:outer_coordinates_v}
\end{align}
The range of both $U_+$ and $V_+$ is $-\infty$ to $\infty$. However, we can define new coordinates $\widetilde{U}_+=\tanh U_+$ and $\widetilde{V}_+=\tanh{V_+}$, which each range from $-1$ to $1$. This allows us to visualize the RN spacetime by drawing a Penrose diagram (see figure~\ref{fig:inner_outer_coordinates}).

Similarly, the set of coordinates regular at the inner horizon is
\begin{align}
U_-&=\begin{cases}
\rme^{-\kappa_- (r_*-t)}& \text{II, III}'\\
-\rme^{-\kappa_- (r_*-t)}& \text{II}',\text{ III}
\end{cases}
\end{align}
\begin{align}
 V_-&= \begin{cases}
\rme^{-\kappa_-(r_*+t)} & \text{II, III}\\
-\rme^{-\kappa_-(r_*+t)} & \text{II}',\text{ III}'
\end{cases}.
\end{align}
The range of $U_-$ and $V_-$ is not the whole real line, since the lower limit $r=0$ is a curve in $U_-V_-$-space, beyond which $U_-$ and $V_-$ are not defined. Nevertheless, we can still define $\widetilde{U}_-=\tanh U_-$ and $\widetilde{V}_-=\tanh{V_-}$. The Penrose diagram for the RN spacetime in these coordinates is shown in figure~\ref{fig:inner_outer_coordinates}.

\begin{figure}
\centering
\includegraphics[scale=.9]{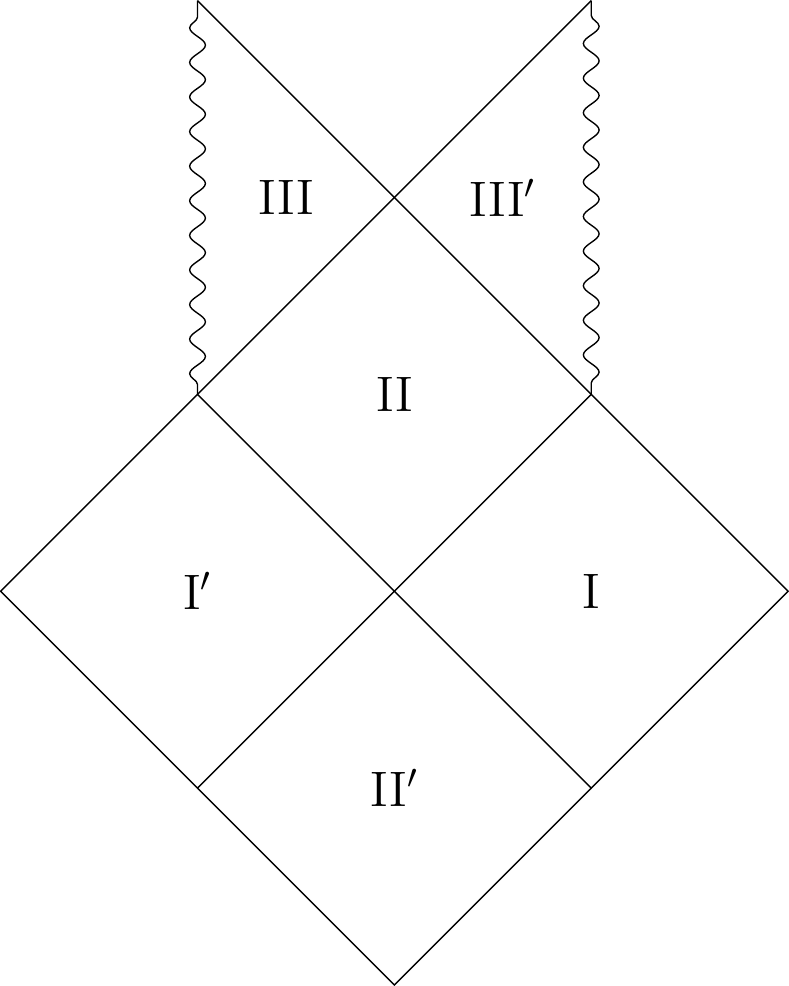}
\caption{The Penrose diagram for regions in an analytically extended RN spacetime.}
\label{fig:complete_reissner}
\end{figure}

\begin{figure}
\centering
\includegraphics[scale=.328]{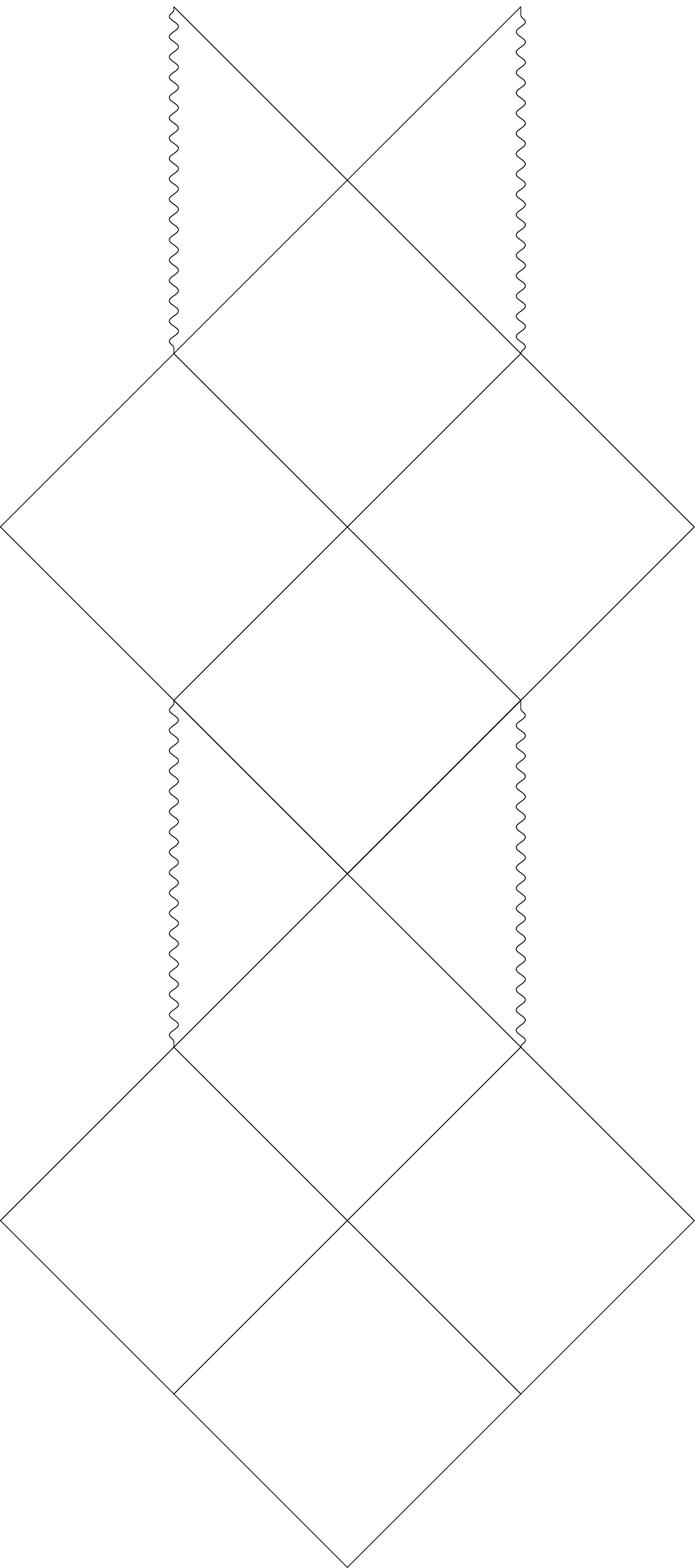}
\caption{A section of the Penrose diagram for a maximal analytic extension of the RN spacetime, which proceeds infinitely in both directions.}
\label{fig:reissner_chain}
\end{figure}

Using these coordinates, we can construct an extended RN spacetime with arbitrarily many spacetime regions. Shown in figure~\ref{fig:complete_reissner} is what we call a single extended RN spacetime, which contains two asymptotically flat regions or universes. You may notice that region II is covered by both the $U_+V_+$- and $U_-V_-$-coordinates. Shown in figure~\ref{fig:reissner_chain} is a section of a maximal analytic extension of the RN spacetime.

\section{Solving the Klein-Gordon Equation}
\label{sec:klein_gordon}
Now that we have two convenient coordinate systems to cover the RN manifold, we can work towards defining a quantum field theory. The goal of this paper is to solve the massless Klein-Gordon equation, which is the classical field equation for a massless scalar field, leaving the definition of the corresponding field operators to future work. While we do not have analytic solutions, we can characterize a basis of two linearly independent solutions for each spacetime region in terms of scattering coefficients. We can then use these bases to construct global solutions for an analytic extension of RN.

\subsection{Separation of Variables}
\label{sec:sep_var}
In this section we use the method of separation of variables on the Klein-Gordon equation, showing that the limiting behavior of the solution is that of a spherical wave. The Klein-Gordon equation in a general spacetime metric $g$ for a massless scalar field $\varphi$ is
\begin{align}
\Box\varphi\equiv \frac{1}{\sqrt{-g}}\partial_\mu\big(\sqrt{-g}g^{\mu\nu}\partial_\nu \varphi\big)=0,
\label{eq:klein_gordon}
\end{align}
where $g=\det g_{\mu\nu}$. For the RN spacetime in spherical coordinates,
\begin{align}
g=-r^4 \sin^2 \theta,
\end{align}
and so using the RN spacetime in~\eqref{eq:klein_gordon},
\begin{equation}
\Box \varphi=-\frac{1}{f}\partial_t{}^2 \varphi +\frac{1}{r^2}\partial_r\big(r^2 f \partial_r \varphi\big)+\frac{1}{r^2 \sin \theta}\partial_\theta \big( \sin\theta \partial_\theta \varphi\big)+\frac{1}{r^2\sin^2\theta}\partial_\phi{}^2 \varphi.
\end{equation}
Using the method of separation of variables, we define $\varphi(t,r,\theta,\phi)=T(t)R(r)Y(\theta,\phi)$. Applying this to the Klein-Gordon equation,
\begin{equation}
0=-\frac{1}{T}\frac{r^2}{f}\partial_t{}^2 T +\frac{1}{R}\partial_r\big(r^2 f \partial_r R\big)+\frac{1}{Y}\Big(\frac{1}{\sin \theta}\partial_\theta\big(\sin\theta\partial_\theta Y\big)+ \frac{1}{\sin^2 \theta }\partial_\phi{}^2Y\Big)
\label{eq:separated_kg}
\end{equation}
This successfully separates the $t$ and $r$ dependence from the angular dependence. 

Equation~\eqref{eq:separated_kg} implies that the angular part must be equal to a constant, which we call $-l(l+1)$:
\begin{equation}
-l(l+1)=(1/Y_l)\Big(\frac{1}{\sin \theta}\partial_\theta\big(\sin\theta\partial_\theta Y_l\big)+ \frac{1}{\sin^2 \theta }\partial_\phi{}^2Y_l\Big)
\label{eq:ang_part}
\end{equation}
We can perform a further separation of variables, which yields the familiar spherical harmonics:
\begin{equation}
Y_l(\theta,\phi)=\sum_{m=-l}^l \alpha_{lm} Y^m_l(\theta,\phi)=\sum_{m=-l}^l \alpha_m \sqrt{\frac{(2l+1)(l-m)!}{4\pi(l+m)!}}P_l^m(\cos \theta) \rme^{\rmi m\phi},
\end{equation}
where $Y_l^m$ are the spherical harmonics, $P_l^m$ are the associated Legendre polynomials, and the $\alpha_{lm}$ are constants.

Substituting~\eqref{eq:ang_part} into~\eqref{eq:separated_kg}, we have
\begin{align}
-\frac{1}{T}\partial_t{}^2 T +\frac{1}{R}\frac{f}{r^2}\partial_r\big(r^2 f \partial_r R\big)-\frac{l(l+1)}{r^2}f=0.
\label{eq:part_separated}
\end{align}
This separates the time part from the radial part, implying that the time part must be equal to a constant, which we call $\omega^2$:
\begin{align}
-\frac{1}{T}\partial_t{}^2 T=\omega^2.
\label{eq:time_part}
\end{align}
We use the solution
\begin{align}
T(t)=\rme^{- \rmi \omega t}.
\end{align}

So far, we've found the angular and time parts of a solution to the Klein-Gordon equation of the form $\varphi=\rme^{- \rmi \omega t}R(r)Y^m_l(\theta,\phi)$. To find $R(r)$, we substitute~\eqref{eq:time_part} into~\eqref{eq:part_separated} to obtain an ordinary differential equation for $R(r)$:
\begin{align}
\frac{1}{R}\frac{f}{r^2}\partial_r\big(r^2 f \partial_r R\big)-\frac{l(l+1)}{r^2}f+\omega^2=0,
\label{eq:radial_orig}
\end{align}
where $f(r)$ is given in~\eqref{eq:orig_f}. While not in its canonical form, this is an example of a confluent Heun equation~\citep{ronveaux1995, slavianov2000}. Note that, as an equation in $r$,~\eqref{eq:radial_orig} has regular singular points at $r=r_-$ and $r=r_+$, and an irregular singular point of rank 1 at $r=\infty$.

To simplify~\eqref{eq:radial_orig} further, we introduce a new function
\begin{align}
A(r)=\frac{r}{r_0}R(r).
\end{align}
The $r_0$ is an arbitrarily chosen constant with dimensions of length in order to preserve units. For future convenience, we set $r_0=r_+$ for the remainder of the paper. In terms of $A(r)$,~\eqref{eq:radial_orig} becomes
\begin{align}
f\partial_r\big(f\partial_rA\big)+\Big(-\frac{f}{r}(\partial_r f)-\frac{l(l+1)}{r^2}f+\omega^2\Big)A=0.
\label{eq:radial_r}
\end{align}
Now we can change variables to the tortoise coordinate $r_*$ (see~\eqref{eq:tortoise}), and we finally have
\begin{align}
\frac{\rmd^2 A(r_*)}{\rmd r_*{}^2}-(V_\text{eff}(r)-\omega^2)A(r_*)=0,
\label{eq:radial}
\end{align}
where
\begin{align}
V_\text{eff}=\frac{f}{r}(\partial_r f)+\frac{l(l+1)}{r^2}f
\end{align}
is the effective potential.

While we cannot write~\eqref{eq:radial} explicitly in terms of $r_*$, we have reduced the radial equation to a time-independent Schr\"odinger-type equation. In particular, $V_\text{eff}\rightarrow 0$ as $r\rightarrow r_\pm,\infty$. This means that the two linearly independent solutions near the boundaries (singular points) $r=r_\pm$ and $r\rightarrow\infty$ are
\begin{align}
R(r_*)\rightarrow \begin{cases}\frac{r_+}{r}\rme^{+ \rmi \omega r_*}\\\frac{r_+}{r}\rme^{-\rmi\omega r_*}\end{cases} \text{ as } r\rightarrow r_+, r_-, \infty.
\end{align}
Even though we have not analytically solved the Klein-Gordon equation, we have enough information to single out solutions with unique limiting behavior. Since $r_*$ diverges at all three of these limits according to~\eqref{eq:rstar_limits}, the radial solution infinitely oscillates near each of these boundaries, approaching a constant amplitude as $r\rightarrow r_\pm$, and decreasing in amplitude as $r\rightarrow\infty$. At this point, we can begin characterizing solutions in each spacetime region in terms of scattering coefficients.

\subsection{Solutions in Each Spacetime Region}

In Section~\ref{sec:sep_var}, we showed that near the horizons and at spatial infinity, the solution to the Klein-Gordon equation is a spherical wave.  Since~\eqref{eq:radial} is a homogeneous second-order ordinary differential equation, a basis for the solution space has two solutions for each spacetime region I, II, and III. In this section, we specify the boundary conditions for each solution in each spacetime region.

\subsubsection{Regions {\normalfont I} and {\normalfont I}\ensuremath{'}}
Regions I and I$'$ are identical except for a reversal in the flow of the Killing vector field $\partial_t$. This results in a simple 180 degree rotation in the Penrose diagram, so the same basis of solutions will work for both regions. For the first basis solution, we specify the boundary condition $R_\text{I}\rightarrow T \rme^{-\rmi\omega r_*}$ as $r\rightarrow r_+$. This means that the $r_+$ horizon emits no waves. $T$ is chosen so that when this boundary condition propagates to infinity, 
\begin{align}
R_\text{I}=\begin{cases} \frac{r_+}{r}T \rme^{-\rmi\omega r_*}& r\rightarrow r_+\\
\frac{r_+}{r}(\rme^{-\rmi\omega r_*}+R\rme^{\rmi\omega r_*})& r\rightarrow \infty\end{cases}\label{eq:I_basis_1}.
\end{align}
The choice of $1$ as the coefficient of $e^{-i\omega r_*}$ inside the bracket provides the second boundary condition and a scale for this solution. 

\begin{figure}
\begin{minipage}{.48\textwidth}
\centering
\includegraphics[scale=1.2]{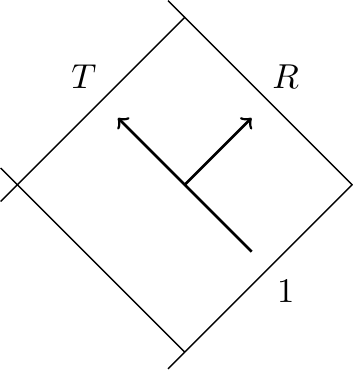}

\vspace{.5cm}

$R_\text{I}=\begin{cases} \frac{r_+}{r}T \rme^{-\rmi\omega r_*}& r\rightarrow r_+\\
\frac{r_+}{r}(\rme^{-\rmi\omega r_*}+R\rme^{\rmi\omega r_*})& r\rightarrow \infty\end{cases}$
\end{minipage}
\begin{minipage}{.48\textwidth}
\centering
\includegraphics[scale=1.2]{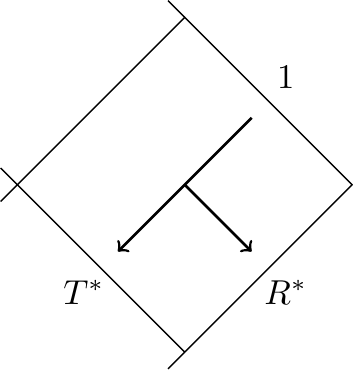}

\vspace{.5cm}

$R_\text{I}{}^*=\begin{cases}\hspace{4em} \frac{r_+}{r}T^* \rme^{\rmi\omega r_*}& r\rightarrow r_+\\
\frac{r_+}{r}(R^*\rme^{-\rmi\omega r_*}+\rme^{\rmi\omega r_*})& r\rightarrow \infty\end{cases}$
\end{minipage}
\caption{A basis of solutions for region I.}
\label{fig:I_basis}
\end{figure}

We can see from~\eqref{eq:radial}, which is real when $r$ and $\omega$ are real, that $R_\text{I}{}^*$ is another, linearly independent solution, which we take to be our second basis solution (unless $T=0$):
\begin{align}
R_\text{I}{}^*=\begin{cases}\hspace{4em} \frac{r_+}{r}T^* \rme^{\rmi\omega r_*}& r\rightarrow r_+\\
\frac{r_+}{r}(R^*\rme^{-\rmi\omega r_*}+\rme^{\rmi\omega r_*})& r\rightarrow \infty\end{cases}.\label{eq:I_basis_2}
\end{align}
This solution has no waves entering the future $r_+$ horizon, and can be viewed as  a time-reversed $R_\text{I}$. Recalling that the time part is $T(t)=\rme^{-\rmi\omega t}$, we can distinguish the ingoing and outgoing parts and thus visualize these solutions as in figure~\ref{fig:I_basis}.

We can now use the Wronskian to find a relation between these scattering coefficients. Since~\eqref{eq:radial} has no first-order derivatives, the Wronskian $W(\tfrac{r}{r_+}R_\text{I},\tfrac{r}{r_+}R_\text{I}{}^*)$ is constant in $r_*$. We calculate
\begin{align*}
W(\tfrac{r}{r_+}R_\text{I},\tfrac{r}{r_+}R_\text{I}{}^*)\rvert_{r\rightarrow\infty}&=2\rmi\omega(1-\lvert R \rvert^2),\\ W(\tfrac{r}{r_+}R_\text{I},\tfrac{r}{r_+}R_\text{I}{}^*)\rvert_{r\rightarrow r_+}&=2\rmi\omega \lvert T \rvert^2.
\end{align*}
This implies that
\begin{align}
\lvert T\rvert^2 + \lvert R \rvert^2 = 1,
\end{align}
which we interpret as conservation of energy.

\subsubsection{Regions {\normalfont II} and {\normalfont II}\ensuremath{'}}

\begin{figure}
\begin{minipage}{.48\textwidth}
\centering
\includegraphics[scale=1.2]{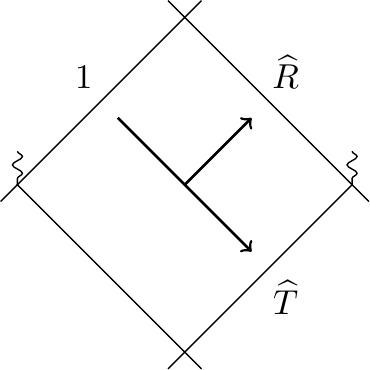}

\vspace{.5cm}

$R_\text{II}=\begin{cases}\frac{r_+}{r}(\rme^{-\rmi\omega r_*}+\widehat{R}\rme^{\rmi\omega r_*})& r\rightarrow r_-\\
\frac{r_+}{r}\widehat{T} \rme^{-\rmi\omega r_*}
& r\rightarrow r_+\end{cases}$
\end{minipage}
\begin{minipage}{.48\textwidth}
\centering
\includegraphics[scale=1.2]{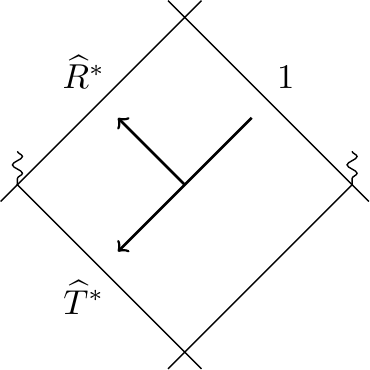}

\vspace{.5cm}

$R_\text{II}{}^*=\begin{cases}\frac{r_+}{r}(\widehat{R}^*\rme^{-\rmi\omega r_*}+\rme^{\rmi\omega r_*})& r\rightarrow r_-\\
\hspace{4em} \frac{r_+}{r}\widehat{T}^* \rme^{\rmi\omega r_*}& r\rightarrow r_+\end{cases}$
\end{minipage}
\caption{A basis of solutions for region II.}
\label{fig:II_basis}
\end{figure}

Similarly to Region I, we specify the boundary condition $R_\text{II}\rightarrow \widehat{T}\rme^{-\rmi\omega r_*}$ as $r\rightarrow r_+$ and let the solution propagate to $r=r_-$. This solution along with its complex conjugate supply the basis of solutions
\begin{align}
R_\text{II}&=\begin{cases}\frac{r_+}{r}(\rme^{-\rmi\omega r_*}+\widehat{R}\rme^{\rmi\omega r_*})& r\rightarrow r_-\\
\frac{r_+}{r}\widehat{T} \rme^{-\rmi\omega r_*}
& r\rightarrow r_+\end{cases}\label{eq:II_basis_1}\\
R_\text{II}{}^*&=\begin{cases}\frac{r_+}{r}(\widehat{R}^*\rme^{-\rmi\omega r_*}+\rme^{\rmi\omega r_*})& r\rightarrow r_-\\
\hspace{4em} \frac{r_+}{r}\widehat{T}^* \rme^{\rmi\omega r_*}& r\rightarrow r_+\end{cases}\label{eq:II_basis_2}
\end{align}
for region II, as shown in figure~\ref{fig:II_basis}.

Again, we can use the Wronskian to find a relation between the scattering coefficients:
\begin{align*}
W(\tfrac{r}{r_+}R_\text{II},\tfrac{r}{r_+}R_\text{II}{}^*)\rvert_{r\rightarrow r_+}&=2\rmi\omega\lvert \widehat{T} \rvert^2,\\
W(\tfrac{r}{r_+}R_\text{II},\tfrac{r}{r_+}R_\text{II}{}^*)\rvert_{r\rightarrow r_-}&=2\rmi\omega(1-\lvert \widehat{R}\rvert^2),
\end{align*}
and thus
\begin{align}
\lvert \widehat{T}\rvert^2 + \lvert \widehat{R}\rvert^2=1.
\label{eq:wronskian_II}
\end{align}
Note that this consequence of the Wronskian relation depends on the choice of basis in region II and may not be standard. For example, in~\citep{kehle2019}, the authors choose a basis from
\begin{align}
    R_\text{II}'&=\begin{cases}\frac{r_+}{r}(\mathfrak{T}\rme^{-\rmi\omega r_*}+\mathfrak{R}\rme^{\rmi\omega r_*})& r\rightarrow r_-\\
\frac{r_+}{r} \rme^{-\rmi\omega r_*}
\end{cases}
\end{align}
and its complex conjugate. This choice is related to ours by
\begin{align}
\mathfrak{T}=1/\widehat{T},\qquad \mathfrak{R}=\widehat{T}/\widehat{R}.    
\end{align}
The authors in~\citep{kehle2019} then obtain
\begin{align}
    |\mathfrak{T}|^2-|\mathfrak{R}|^2=1.
\end{align}

\subsubsection{Regions {\normalfont III} and {\normalfont III}\ensuremath{'}}

As we've already determined, $R(r)\rightarrow \rme^{\pm \rmi \omega r_*}$ as $r\rightarrow r_-$, but it is not immediately clear from~\eqref{eq:radial} how $R(r)$ behaves as $r\rightarrow 0$, since $V_\text{eff}$ diverges there and the definition of $A(r)$ breaks down. In fact, we have to return to the radial equation in the form of~\eqref{eq:radial_orig} to test the regularity or singularity of the scalar wave equation at $r=0$. We can rewrite the equation in the form
\begin{align}
\frac{\rmd^2 R}{\rmd r^2}+\frac{\partial_r (r^2f)}{r^2f}\frac{\rmd R}{\rmd r}+\Big(-\frac{l(l+1)}{r^2f}+\frac{\omega}{f^2}\Big) R=0.
\end{align}
Now, in the limit of $r\rightarrow 0$,
\begin{align*}
\frac{\partial_r (r^2f)}{r^2f}\rightarrow -\frac{2M}{Q^2},\qquad
-\frac{l(l+1)}{r^2f}+\frac{\omega}{f^2}\rightarrow -\frac{l(l+1)}{Q^2}.
\end{align*}
Since both of these limits are finite, $r=0$ is a regular point of the radial equation, despite being a geometric singularity. Taking a Taylor series about this point, the two linearly independent behaviors as $r\rightarrow 0$ are $R\rightarrow 1,r/r_1$. Here $r_1$ is arbitrarily chosen to preserve units, and we set $r_1=r_-$ for future convenience. We use these as boundary conditions to obtain our basis of solutions for region III.

\begin{figure}
\begin{minipage}{.5\textwidth}
\centering
\includegraphics[scale=1.2]{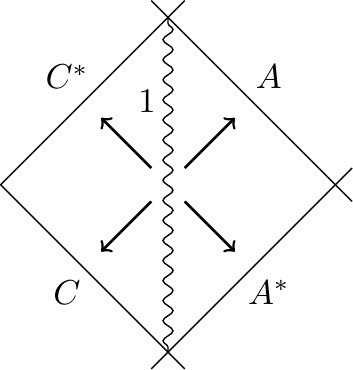}

\vspace{.5cm}

$R_{\text{III},1}=\begin{cases}
\frac{r_+}{r}(C^*e^{-i\omega r_*}+Ce^{i\omega r_*})& r\rightarrow -\infty \\
1&r\rightarrow 0\\
\frac{r_+}{r}(A^*e^{-i\omega r_*}+Ae^{i\omega r_*})& r\rightarrow r_-\end{cases}$
\end{minipage}
\begin{minipage}{.5\textwidth}
\centering
\includegraphics[scale=1.2]{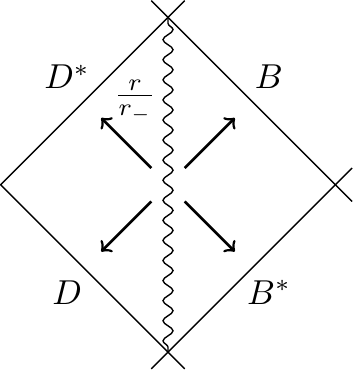}

\vspace{.5cm}

$R_{\text{III},r}=\begin{cases}
\frac{r_+}{r}(D^*\rme^{-\rmi\omega r_*}+D\rme^{\rmi\omega r_*})& r\rightarrow -\infty\\
\frac{r}{r_-}&r\rightarrow 0\\
\frac{r_+}{r}(B^*\rme^{-\rmi\omega r_*}+B\rme^{\rmi\omega r_*})& r\rightarrow r_-\end{cases}$
\end{minipage}
\caption{A basis of solutions for region III.}
\label{fig:III_basis}
\end{figure}

In addition, since $r=0$ is a regular point of the field equation, if necessary we can append a spacetime with negative $r$. Where $r<0$ there are no horizons, so the spacetime has a naked singularity~\cite{Giveon:2004yg}. The boundary conditions  ``scatter'' from the singularity, and we obtain the basis
\begin{align}
R_{\text{III},1}&=\begin{cases}
\frac{r_+}{r}(C^*e^{-i\omega r_*}+Ce^{i\omega r_*})& r\rightarrow -\infty \\
1&r\rightarrow 0\\
\frac{r_+}{r}(A^*e^{-i\omega r_*}+Ae^{i\omega r_*})& r\rightarrow r_-\end{cases}\label{eq:III_basis1_1}\\
R_{\text{III},r}&=\begin{cases}
\frac{r_+}{r}(D^*\rme^{-\rmi\omega r_*}+D\rme^{\rmi\omega r_*})& r\rightarrow -\infty\\
\frac{r}{r_-}&r\rightarrow 0\\
\frac{r_+}{r}(B^*\rme^{-\rmi\omega r_*}+B\rme^{\rmi\omega r_*})& r\rightarrow r_-\end{cases}\label{eq:III_basis1_2}
\end{align}
for region III, as shown in figure~\ref{fig:III_basis}.
We can use the Wronskian along with the definition~\eqref{eq:tortoise} to find a relation between $A$, $B$, $C$, and $D$:
\begin{align*}
W(\tfrac{r}{r_+}R_{\text{III},1},\tfrac{r}{r_+}R_{\text{III},r})\rvert_{r\rightarrow r_-}
&=2\rmi\omega(A^* B- A B^*),\\
W(\tfrac{r}{r_+}R_{\text{III},1},\tfrac{r}{r_+}R_{\text{III},r})\rvert_{r\rightarrow -\infty}&=2\rmi\omega(C^* D- C D^*),\\
W(\tfrac{r}{r_+}R_{\text{III},1},\tfrac{r}{r_+}R_{\text{III},r})\rvert_{r\rightarrow 0}&=\frac{(r-r_+)(r-r_-)}{r_+{}^2r_-}\rightarrow \frac{1}{r_+},
\end{align*}
and so we have
\begin{align}
AB^*-A^*B=CD^*-C^*D=\frac{\rmi}{2r_+\omega }.
\end{align}

\begin{figure}
\begin{minipage}{.5\textwidth}
\centering
\includegraphics[scale=1.2]{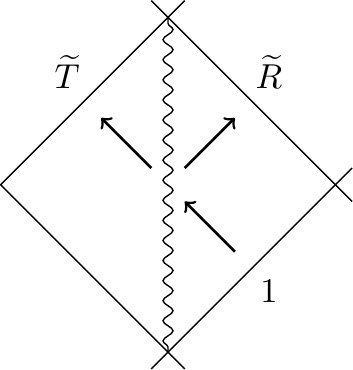}

\vspace{.5cm}

$R_{\text{III}}=\begin{cases}
\frac{r_+}{r}\widetilde{T}\rme^{-\rmi\omega r_*}& r\rightarrow -\infty \\
\frac{r_+}{r}(\rme^{-\rmi\omega r_*}+\widetilde{R}\rme^{\rmi\omega r_*})& r\rightarrow r_-\end{cases}$
\end{minipage}
\begin{minipage}{.5\textwidth}
\centering
\includegraphics[scale=1.2]{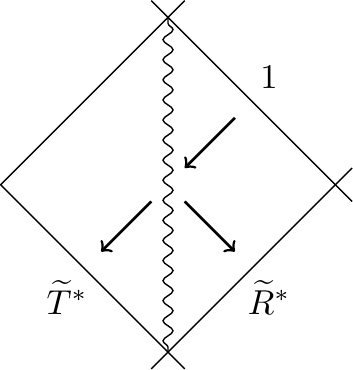}

\vspace{.5cm}

$R_{\text{III}}{}^*=\begin{cases}
\frac{r_+}{r}\widetilde{T}^*\rme^{\rmi\omega r_*}& r\rightarrow -\infty\\
\frac{r_+}{r}(\widetilde{R}^*\rme^{-\rmi\omega r_*}+\rme^{\rmi\omega r_*})& r\rightarrow r_-\end{cases}$
\end{minipage}
\caption{An alternative basis of solutions for region III.}
\label{fig:III_basis_2}
\end{figure}

If we are unconcerned with the behavior at $r=0$, but would still like to append a geometry with negative $r$, it might be convenient to introduce an alternative basis for region III, which specifies boundary conditions at $r=r_-$ and $r=-\infty$ instead of at $r=0$. One such basis is
\begin{align}
    R_{\text{III}}&=\begin{cases}
\frac{r_+}{r}\widetilde{T}\rme^{-\rmi\omega r_*}& r\rightarrow -\infty \\
\frac{r_+}{r}(\rme^{-\rmi\omega r_*}+\widetilde{R}\rme^{\rmi\omega r_*})& r\rightarrow r_-\end{cases}\label{eq:III_basis2_1}\\
R_{\text{III}}{}^*&=\begin{cases}
\frac{r_+}{r}\widetilde{T}^*\rme^{\rmi\omega r_*}& r\rightarrow -\infty\\
\frac{r_+}{r}(\widetilde{R}^*\rme^{-\rmi\omega r_*}+\rme^{\rmi\omega r_*})& r\rightarrow r_-\end{cases}\label{eq:III_basis2_2}
\end{align}
as shown in figure~\ref{fig:III_basis_2}. Computing the Wronskians, we find the relationships between the scattering coefficients $\widetilde{T}$, $\widetilde{R}$, $A$, $B$, $C$, and $D$:
\begin{align*}
W(\tfrac{r}{r_+}R_{\text{III}},\tfrac{r}{r_+}R_{\text{III}}{}^*)|_{r\rightarrow -\infty}&=2\rmi\omega|\widetilde{T}|^2,\\
W(\tfrac{r}{r_+}R_{\text{III}},\tfrac{r}{r_+}R_{\text{III}}{}^*)|_{r\rightarrow -r_-}&=2\rmi\omega(1-|\widetilde{R}|^2)\\
W(\tfrac{r}{r_+}R_{\text{III}},\tfrac{r}{r_+}R_{\text{III},1})|_{r\rightarrow -\infty}&=2\rmi\omega\widetilde{T}C\\
W(\tfrac{r}{r_+}R_{\text{III}},\tfrac{r}{r_+}R_{\text{III},1})|_{r\rightarrow -r_-}&=2\rmi\omega(A-\widetilde{R}A^*)\\
W(\tfrac{r}{r_+}R_{\text{III}},\tfrac{r}{r_+}R_{\text{III},r})|_{r\rightarrow -\infty}&=2\rmi\omega\widetilde{T}D\\
W(\tfrac{r}{r_+}R_{\text{III}},\tfrac{r}{r_+}R_{\text{III},r})|_{r\rightarrow -r_-}&=2\rmi\omega(B-\widetilde{R}B^*).
\end{align*}
Solving the above systems of equations, we derive the following relations:
\begin{align}
1&=|\widetilde{T}|^2+|\widetilde{R}|^2,\\
\widetilde{T}&=\frac{AB^*-A^*B}{B^*C-A^*D},\\
\widetilde{R}&=\frac{BC-AD}{B^* C - A^* D}.
\end{align}

We now have a basis of solutions for every spacetime region in an arbitrarily extended RN manifold. In the next section, we demonstrate how to put these together into a global solution.

\subsection{Building Global Solutions}

In the previous section we constructed a basis of solutions to the Klein-Gordon equation in a single spacetime region. We will now use this basis to build a global basis of solutions for the manifold shown in figure~\ref{fig:complete_reissner} as well as a maximally extended RN spacetime shown in figure~\ref{fig:reissner_chain}.

\subsubsection{An Extended Reissner-Nordstr\"om Solution}

\begin{figure*}[ht]
\centering
\includegraphics[scale=1]{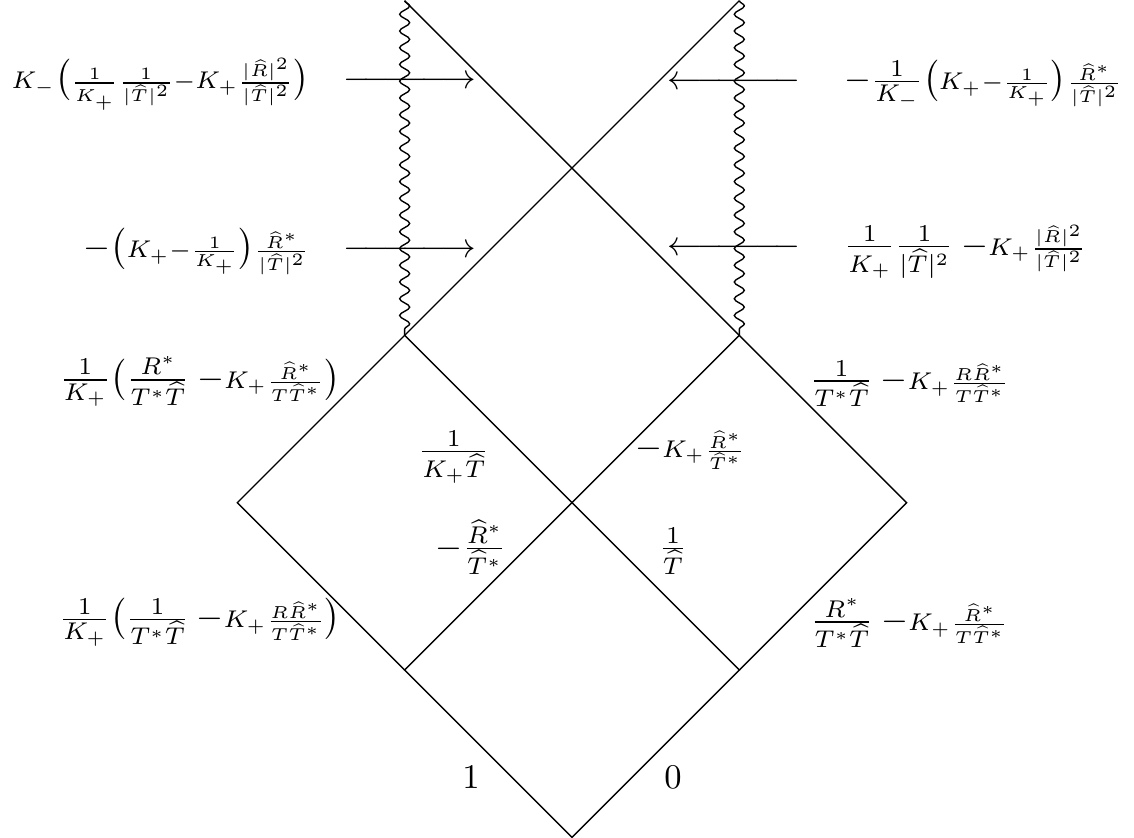}
\caption{A representation of a global, positive frequency solution for an extended RN geometry.}
\label{fig:extended_5}
\end{figure*}

We will now construct a global solution of positive frequency for the extended RN spacetime shown in figure~\ref{fig:complete_reissner}. The result is shown in figure~\ref{fig:extended_5}, but the first step to obtain this solution is to specify a boundary condition within a single spacetime region. It will be convenient to specify  radiation at only one of the inner horizon boundaries in region II$'$. In region II$'$, this solution is a linear combination of $R_\text{II}$ and $R_{\text{II}}{}^*$:
\begin{align}
R(r)\rvert_{\text{II}'}=-\frac{\widehat{R}^*}{\lvert\widehat{T}\rvert^2}R_\text{II}+\frac{1}{\lvert \widehat{T}\rvert^2 }R_\text{II}{}^*.
\end{align}

The propogation of a solution across the outer horizon (i.e. across the $U_+$- or $V_+$-axis) requires consideration of the analyticity of the solution on the horizon. There are two solutions analytical at the horizon for a given solution in region I. At the point $(-\widetilde{U}_+,-\widetilde{V}_+)$ in region I$'$, the value of the so-called ``positive frequency'' solution is $1/K_+$  times the value at the point  $(\widetilde{U}_+,\widetilde{V}_+)$ in region I, where
\begin{align}
K_\pm=\rme^{\pi\omega/\kappa_\pm}.
\end{align}
For negative frequency solutions, the multiplicative factor is $K_+$.  Similarly at the inner horizon, at the point $( \widetilde{U}_-,-\widetilde{V}_-)$ in region III$'$, the value of the positive frequency solution is $1/K_-$ times the value at the point $( -\widetilde{U}_-,\widetilde{V}_-)$. For negative frequency solutions, the multiplicative factor is $K_-$. See~\ref{sec:pos_neg_freq} for the definitions of the positive and negative frequency solutions.

Proceeding by choosing positive frequency at both horizons and finding the correct linear combination for region II, we find the propagation in figure~\ref{fig:extended_5}. The boundaries at $r=0$ and $r=-\infty$ can be found by taking the correct linear combinations of~\eqref{eq:III_basis1_1} and~\eqref{eq:III_basis1_2} (or~\eqref{eq:III_basis2_1} and ~\eqref{eq:III_basis2_2} for $r=-\infty$), but the expressions are complicated and irrelevant to our analysis of maximally extended solutions below. At this point, we have constructed a global solution to a single extended RN geometry from two boundary conditions and two frequency choices (one at each horizon).

\subsubsection{A Maximally Extended Reissner-Nordstr\"om Solution}
\label{sec:extended_solution}

One could continue the process outlined in the previous section in order to construct a basis of solutions in a maximally extended RN geometry (see figure~\ref{fig:reissner_chain}), but beyond a single extended RN geometry, the expressions get long and complicated. Instead, we will formulate the propagation of the scalar field through single RN geometries as an eigenvalue problem in the two-dimensional space of solutions with positive frequency at every horizon. We can then determine the amplitude of the scalar field arbitrarily far from the boundary conditions.

\begin{figure*}[t]
\begin{minipage}{\textwidth}
\centering
\hspace{-1.3em}
\includegraphics[trim=0 0 0 0,clip,scale=1]{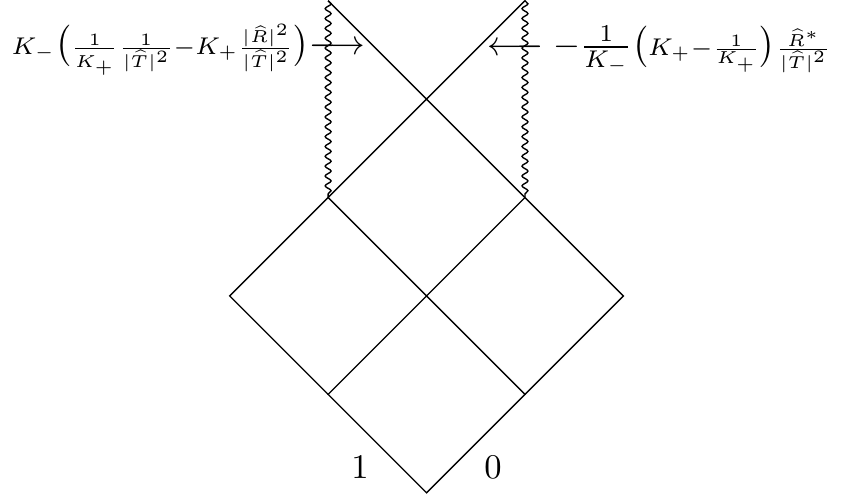}

$\phi_1$
\end{minipage}
\begin{minipage}{\textwidth}
\centering
\includegraphics[trim=-12 0 0 0,clip,scale=1]{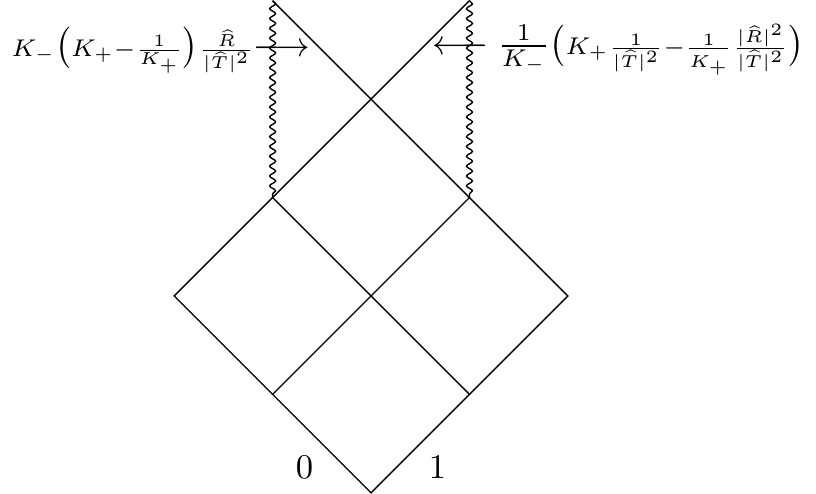}

$\phi_2$
\end{minipage}
\caption{Any positive-frequency solution to a single extended RN geometry can be written in the basis of the solutions $\phi_1$ and $\phi_2$ above.}
\label{fig:pos_basis}
\end{figure*}

After fixing positive frequency at all horizons, there are two linearly independent global solutions for the entire maximally extended RN geometry. We choose the two solutions $\phi_1$ and $\phi_2$ shown in figure~\ref{fig:pos_basis}, whose domains are now the maximally extended RN manifold. The set $\{\phi_1,\phi_2\}$ forms a basis for the linear space of globally positive frequency solutions, so we can write any solution $\phi=\alpha_0\phi_1+\beta_0 \phi_2$, where $\alpha_0$ and $\beta_0$ are the boundary conditions for $\phi$ on the inner horizon in region II$'$. While this choice is arbitrary, equivalent bases can be obtained by setting the boundary data elsewhere.

Using the amplitudes of $\phi_1$ and $\phi_2$ shown in figure~\ref{fig:pos_basis}, we find that on the  inner horizon of region II of the $n$\textsuperscript{th} extended RN geometry (proceeding to the future), the amplitudes $\alpha_n$ and $\beta_n$ are
\begin{equation}
\begin{pmatrix}
\alpha_n \\\beta_n
\end{pmatrix}=M_{++}{}^n
\begin{pmatrix}
\alpha_0\\\beta_0
\end{pmatrix},
\end{equation}
where
\begin{equation}
M_{++}\equiv \begin{pmatrix}
\scriptstyle{K_-}\big(\frac{1}{K_+}-\scriptstyle{K_+}\lvert\widehat{R}\rvert^2\big)\frac{1}{\lvert \widehat{T}\rvert^2} &&
\scriptstyle{K_-}\big(\scriptstyle{K_+}-\frac{1}{K_+}\big)\frac{\widehat{R}}{\lvert \widehat{T}\rvert^2}\\
-\frac{1}{K_-}\big(\scriptstyle{K_+}-\frac{1}{K_+}\big)\frac{\widehat{R}^*}{\lvert \widehat{T}\rvert^2}&&
\frac{1}{K_-}\big(\scriptstyle{K_+}-\frac{1}{K_+}\lvert \widehat{R}\rvert^2\big)\frac{1}{\lvert \widehat{T}\rvert^2}
\end{pmatrix}.
\end{equation}
We note in passing that the uniform boundedness of the scattering data as defined in~\citep{kehle2019} applies directly in the components of the matrix $M_{++}$.

The evolution of the field depends entirely upon the properties of the matrix $M_{++}$ (where the two positive signs stand for positive frequency at both horizons). Using the Wronskian relation~\eqref{eq:wronskian_II}, we find
\begin{equation}
\det M_{++}=1
\end{equation}
and
\begin{equation}
\begin{split}
\operatorname{tr}M_{++}&=\frac{K_+}{K_-}+\frac{K_-}{K_+}-\Big(K_+-\frac{1}{K_+}\Big)\Big(K_--\frac{1}{K_-}\Big)\frac{\lvert\widehat{R}\rvert^2}{\lvert \widehat{T}\rvert^2}\\
&=2\cosh4\pi\omega M-4\tfrac{\lvert\widehat{R}\rvert^2}{\lvert \widehat{T}\rvert^2}\sinh(\pi\omega/\kappa_+)\sinh(\pi\omega/\kappa_-)\\
&=\frac{2}{\lvert \widehat{T}\rvert^2}\big( \cosh 4\pi\omega M-\lvert \widehat{R}\rvert^2 \cosh \pi\omega(1/\kappa_+ +1/\kappa_-)\big)\\
&=2\big( |\mathfrak{T}|^2\cosh 4\pi\omega M-\lvert \mathfrak{R}\rvert^2 \cosh \pi\omega(1/\kappa_+ +1/\kappa_-)\big),
\end{split}
\end{equation}
where $\mathfrak{T}$ and $\mathfrak{R}$ refer to the scattering data constructed in~\citep{kehle2019}.  
It will be interesting for later discussion to write the scattering coefficients in terms of the trace:
\begin{align}
\lvert \widehat{T}\rvert^2&=\frac{2\sinh\big(2\pi\omega \tfrac{r_+{}^2}{r_+-r_-}\big)\sinh\big(2\pi\omega \tfrac{r_-{}^2}{r_+-r_-}\big)}{\cosh \big(2\pi\omega\frac{r_+{}^2+r_-{}^2}{r_+-r_-}\big)-\tfrac{1}{2}\operatorname{tr}M_{++}},
\label{eq:scattering_trace_t}\\
\lvert\widehat{R}\rvert^2&=\frac{\cosh \big(2\pi\omega\frac{r_+{}^2-r_-{}^2}{r_+-r_-}\big)-\tfrac{1}{2}\operatorname{tr}M_{++}}{\cosh \big(2\pi\omega\frac{r_+{}^2+r_-{}^2}{r_+-r_-}\big)-\tfrac{1}{2}\operatorname{tr}M_{++}}.
\label{eq:scattering_trace_r}
\end{align}
Since $\det M_{++}>0$, $M_{++}$ is invertible, $M_{++}{}^{-1}$ exists, and we can propagate the field to the past with $M_{++}{}^{-1}$. Using the eigenvalue equation, we can find the eigenvalues $\lambda_\pm$ and the associated eigenvectors $\vec{\lambda}_\pm$:
\begin{align}
\lambda_\pm&= \tfrac{1}{2} \operatorname{tr} M_{++}\pm \sqrt{(\tfrac{1}{2} \operatorname{tr} M_{++})^2-1}\label{eq:matrix_eigenvalues}\\
\vec{\lambda}_\pm &=\begin{pmatrix}
K_-(K_+-\frac{1}{K_+})\frac{\widehat{R}}{\lvert \widehat{T}\rvert^2}\\
\lambda_\pm-K_-(\frac{1}{K_+}-K_+\lvert \widehat{R}\rvert^2)
\end{pmatrix}\propto
\begin{pmatrix}
\lambda_\pm-\frac{1}{K_-}(K_+-\frac{1}{K_+}\lvert \widehat{R}\rvert^2)\\
-\frac{1}{K_-}(K_+-\frac{1}{K_+})\frac{\widehat{R}^*}{\lvert \widehat{T}\rvert^2}
\end{pmatrix}.
\end{align}

Now let us discuss these results. While at this point we know nothing about the $M$, $Q$, and $\omega$ dependence of $\widehat{R}$ and $\widehat{T}$, the eigenvalues depend only on the quantity $\tfrac{1}{2}\operatorname{tr}M_{++}$. If $(\tfrac{1}{2}\operatorname{tr}M_{++})^2< 1$, we have that $\lambda_\pm\notin \mathbb{R}$, $\lambda_+=\lambda_-{}^*$, and $\lvert \lambda_\pm \rvert^2=1$. In this situation we have the desirable property that the amplitude of the solution is non-divergent as one proceeds arbitrarily far to the future or to the past in the maximally extended geometry.  If $(\tfrac{1}{2}\operatorname{tr}M_{++})^2=1$, then $\lambda_\pm\in \mathbb{R}$ and $\lambda_+=\lambda_-=1$. However, we also have that $\vec{\lambda_+}=\vec{\lambda_-}$, which means that $M_{++}$ has become a defective matrix with a double eigenvalue but only a single eigenvector. Thus, the solution associated with this eigenvector will not diverge, but all other solutions will diverge. Lastly, if $(\tfrac{1}{2}\operatorname{tr}M_{++})^2> 1$, then $\lambda_\pm\in \mathbb{R}$, $\lambda_\pm \neq 1$, and $\lambda_+=1/\lambda_-$. In this case, the amplitude of the solutions corresponding to $\lambda_\pm$ will diverge in one direction and converge to zero in the other direction. Any superposition of these two solutions will diverge in both directions.

Clearly the first case is the most desirable as it is the only case where no solutions diverge. A necessary and sufficient condition for a non-divergent solution space is
\begin{align}
\rvert\tfrac{1}{2}\operatorname{tr} M_{++}\lvert<1 &\Longleftrightarrow \quad\frac{\sinh^2(2\pi\omega M)}{\sinh(\pi\omega/\kappa_+)\sinh(\pi\omega/\kappa_-)}< \frac{\lvert \widehat{R}\rvert^2}{\lvert \widehat{T}\rvert^2}< \frac{\cosh^2(2\pi\omega M)}{\sinh(\pi\omega/\kappa_+)\sinh(\pi\omega/\kappa_-)}\\
& \Longleftrightarrow \quad \frac{\sinh 2\pi\omega M}{\sinh   \pi\omega(1/2\kappa_++1/2\kappa_-)}<\lvert \widehat{R}\rvert<\frac{\cosh 2\pi\omega M}{\cosh \pi\omega(1/2\kappa_++1/2\kappa_-)}.
\label{eq:conv_cond}
\end{align}
In order to determine if or when this condition is satisfied, we need to know the dependence of $\widehat{R}$ and $\widehat{T}$ on $M$, $\omega$, $Q$, and $l$. In the next section, we provide analytical and numerical approximations to answer this question.

\section{Scattering Coefficients in Region II}
\label{sec:scattering}

As we saw in the previous section, the global structure of a scalar field on a maximally extended RN manifold is highly dependent on the scattering coefficients for region II, $\widehat{T}$ and $\widehat{R}$. It is possible to write down the exact radial function in terms of Heun-type functions (cf.~\citep{bezerra2014}), but we have found it more useful to approximate the low-frequency limit analytically using the hypergeometric equation, as in~\citep{starobinsky1973}. We also provide a numerically obtained example to demonstrate the limitations of the low-frequency solution.

\subsection{Low Frequency}

In this section, we will obtain expressions for the scattering coefficients for region II in the low frequency limit. To do this, we arrange the radial equation in a form similar to Euler's hypergeometric equation, drop higher order terms in order to match to the hypergeometric equation, and find $\widehat{T}$ and $\widehat{R}$ using the scattering properties of the hypergeometric function.

To begin, we define a re-scaled radial coordinate $x=(r-r_-)/(r_+-r_-)$, in terms of which the radial equation~\eqref{eq:radial_orig} becomes
\begin{multline}
0=\bigg(\frac{\rmd^2}{\rmd x^2}+(r_+-r_-)^2\omega^2
+\frac{1}{x}\Big(\frac{1}{2}+l(l+1)-\frac{r_+{}^4-2r_+{}^3r_--2r_+r_-{}^3+r_-{}^4}{(r_+-r_-)^2}\omega^2+(r_+{}^2-r_-{}^2)\omega^2\Big)
\\+\frac{1}{1-x}\Big(\frac{1}{2}+l(l+1)-\frac{r_+{}^4-2r_+{}^3r_--2r_+r_-{}^3+r_-{}^4}{(r_+-r_-)^2}\omega^2-(r_+{}^2-r_-{}^2)\omega^2\Big)\\
+\frac{1}{x^2}\Big(\frac{1}{4}+\frac{r_-{}^4\omega^2}{(r_+-r_-)^2}\Big)+\frac{1}{(1-x)^2}\Big(\frac{1}{4}+\frac{r_+{}^4\omega^2}{(r_+-r_-)^2}\Big)\bigg)\sqrt{x(1-x)}R(x).
\end{multline}
At this point, we need to drop the terms $(r_+-r_-)^2\omega^2$ and $(r_+{}^2-r_-{}^2)\omega^2 $, which can be accomplished by imposing $M\omega\ll 1$. Doing so, we rearrange to find
\begin{multline}
0=\bigg(\frac{\rmd^2}{\rmd x^2}+\frac{1}{x(1-x)}\Big(\frac{1}{2}+l(l+1)-\frac{r_+{}^4-2r_+{}^3r_--2r_+r_-{}^3+r_-{}^4}{(r_+-r_-)^2}\omega^2\Big)\\
\frac{1}{x^2}\Big(\frac{1}{4}+\frac{r_-{}^4\omega^2}{(r_+-r_-)^2}\Big)+\frac{1}{(1-x)^2}\Big(\frac{1}{4}+\frac{r_+{}^4\omega^2}{(r_+-r_-)^2}\Big)\bigg)\sqrt{x(1-x)}R(x).
\label{eq:hypergeometric_radial}
\end{multline}
The hypergeometric equation for the hypergeometric function $F(a,b,c\,;z)$ can be written in the form
\begin{multline}
0=\bigg(\frac{\rmd^2}{\rmd z^2}+\frac{1}{z(1-z)}\Big(\frac{1}{4}+\Big(\frac{a-b}{2}\Big)^2-\Big(\frac{c-1}{2}\Big)^2 -\Big(\frac{a+b-c}{2}\Big)^2\Big)\\
+\frac{1}{z^2}\Big(\frac{1}{4}-\Big(\frac{c-1}{2}\Big)^2\Big)+\frac{1}{(1-z)^2}\Big(\frac{1}{4}-\Big(\frac{a+b-c}{2}\Big)^2\Big)\bigg) z^{c/2}(1-z)^{(1+a+b-c)/2}F(a,b,c\,;z).
\label{eq:hypergeometric}
\end{multline}
We can now see that the radial equation and the hypergeometric equation are in similar forms. Because the radial solution of interest is
\begin{align}
R_\text{II}\xrightarrow[r\rightarrow r_+]{}&\frac{r_+}{r}\widehat{T} \rme^{-\rmi\omega r_*}=\frac{r_+}{r}\widehat{T}\rme^{-\rmi\omega r}\Big(\frac{r_+-r_-}{r_+}\Big)^{-\frac{\rmi\omega r_+{}^2}{r_+-r_-}}\Big(\frac{r_+-r_-}{r_-}\Big)^{\frac{\rmi\omega r_-{}^2}{r_+-r_-}}(1-x)^{-\frac{\rmi r_+{}^2\omega}{r_+-r_-}}x^{\frac{\rmi r_-{}^2\omega}{r_+-r_-}},
\label{eq:radial_interest}
\end{align}
we find it useful to make the substitution $z=1-x$, which means the radial solution is related to the hypergeometric function by
\begin{align}
R(x)=(1-x)^{(c-1)/2}x^{(a+b-c)/2}F(a,b,c\,; 1-x).
\label{eq:radialToHypergeometric}
\end{align}
With this substitution, in comparing~\eqref{eq:hypergeometric} and~\eqref{eq:hypergeometric_radial} we can make the identifications
\begin{align}
\frac{c-1}{2}&=-\rmi\frac{\omega r_+{}^2}{r_+-r_-}=-\rmi\omega/2\kappa_+
\label{eq:cminus1}\\
\frac{a+b-c}{2}&=+\rmi\frac{\omega r_-{}^2}{r_+-r_-}=+\rmi\omega/2\kappa_-
\label{eq:aplusbminusc}\\
\frac{a-b}{2}&=\frac{1}{2}\sqrt{(2l+1)^2-8(r_+{}^2+r_+r_-+r_-{}^2)\omega^2}.
\label{eq:aminusb}
\end{align}
The sign choices in~\eqref{eq:cminus1} and~\eqref{eq:aplusbminusc} were made to match~\eqref{eq:radialToHypergeometric} to~\eqref{eq:radial_interest}, and the sign choice in~\eqref{eq:aminusb} is arbitrary since it only interchanges $a$ and $b$. Solving the above system of equations, we find
\begin{align}
a&=\frac{1}{2}\Big(1-2\rmi\omega(r_++r_-)+\sqrt{(1+2l)^2-8(r_+{}^2+r_+r_-+r_-{}^2)\omega^2}\Big) \\
b&=\frac{1}{2}\Big(1-2\rmi\omega(r_++r_-)-\sqrt{(1+2l)^2-8(r_+{}^2+r_+r_-+r_-{}^2)\omega^2}\Big) \\
c&=1-\frac{2\rmi r_+{}^2\omega}{r_+-r_-}=1-\rmi\omega/\kappa_+.
\end{align}

Now that we have found $a$, $b$, and $c$, we can use~\eqref{eq:radial_interest} and~\eqref{eq:radialToHypergeometric} to approximate the solution $R_\text{II}$ in~\eqref{eq:II_basis_1}:
\begin{align}
R_\text{II}\approx C\widehat{T}(1-x)^{(c-1)/2}x^{(a+b-c)/2}F(a,b,c\,; 1-x),
\end{align}
where $C$ is a constant phase, which can be calculated from~\eqref{eq:radial_interest}. Then, according to the formula~\citep{abramowitz1964}
\begin{multline}
F(a,b,c\,;z)=\frac{\Gamma(c)\Gamma(c-a-b)}{\Gamma(c-a)\Gamma(c-b)}F(a,b,a+b-c+1\,;1-z)\\+(1-z)^{-(a+b-c)}z^{-(c-1)}\frac{\Gamma(c)\Gamma(a+b-c)}{\Gamma(a)\Gamma(b)} F(1-a,1-b,c-a-b+1\,;1-z),
\end{multline}
where $\Gamma$ is the gamma function, we also have
\begin{multline}
R_\text{II}\approx C\widehat{T}(1-x)^{(c-1)/2}x^{(a+b-c)/2}\frac{\Gamma(c)\Gamma(c-a-b)}{\Gamma(c-a)\Gamma(c-b)}\cdot F(a,b,a+b-c+1\,;x)\\
+C\widehat{T}(1-x)^{-(c-1)/2}x^{-(a+b-c)/2}\frac{\Gamma(c)\Gamma(a+b-c)}{\Gamma(a)\Gamma(b)}F(1-a,1-b,c-a-b+1\,;x).
\end{multline}
We can now calculate the limit as $x\rightarrow 0$:
\begin{align}
R_\text{II}\xrightarrow[r\rightarrow r_-]{\approx}& \,C\widehat{T}x^{\frac{r_-{}^2\omega}{r_+-r_-}}\frac{\Gamma(c)\Gamma(c-a-b)}{\Gamma(c-a)\Gamma(c-b)}+C\widehat{T}x^{-\frac{r_-{}^2\omega}{r_+-r_-}}\frac{\Gamma(c)\Gamma(a+b-c)}{\Gamma(a)\Gamma(b)}\nonumber\\
&=CD\widehat{T}\frac{\Gamma(c)\Gamma(c-a-b)}{\Gamma(c-a)\Gamma(c-b)}e^{-i\omega r_*}+CD^*\widehat{T}\frac{\Gamma(c)\Gamma(a+b-c)}{\Gamma(a)\Gamma(b)}e^{i\omega r_*} \\
R_\text{II}\xrightarrow[r\rightarrow r_-]{}&\, \tfrac{r_+}{r_-}(e^{-\rmi\omega r_*}+\widehat{R}e^{\rmi\omega r_*}),
\end{align}
where $D$ is another constant phase, which can be calculated from~\eqref{eq:radial_interest}. Comparing the last two lines, we conclude
\begin{align}
\frac{1}{\lvert\widehat{T}\rvert}=\frac{r_-}{r_+}\Big\lvert\frac{\Gamma(c)\Gamma(c-a-b)}{\Gamma(c-a)\Gamma(c-b)}\Big\rvert,\qquad
\frac{\lvert\widehat{R}\rvert}{\lvert\widehat{T}\rvert}&=\frac{r_-}{r_+}\Big\lvert\frac{\Gamma(c)\Gamma(a+b-c)}{\Gamma(a)\Gamma(b)}\Big\rvert.
\end{align}
Using the recursive relations $\Gamma(z)=\Gamma(z+1)/z=(z-1)\Gamma(z-1)$ and Euler's reflection formula \mbox{$\Gamma(1-z)\Gamma(z)=\pi/\sin(\pi z)$}, we can find compact expressions for the low-frequency limits of $\lvert \widehat{T}\rvert^2$ and $\lvert \widehat{R}\rvert^2$:
\begin{align}
\lvert \widehat{T}\rvert^2&=\frac{2\sinh\big(2\pi\omega \tfrac{r_+{}^2}{r_+-r_-}\big)\sinh\big(2\pi\omega \tfrac{r_-{}^2}{r_+-r_-}\big)}{\cosh \big(2\pi\omega\frac{r_+{}^2+r_-{}^2}{r_+-r_-}\big)+\cos\big(2\pi\sqrt{(l+\frac{1}{2})^2-2\omega^2(r_+{}^2+r_+r_-+r_-{}^2)}\big)},
\label{eq:lowfreq_t}\\
\lvert\widehat{R}\rvert^2&=\frac{\cosh \big(2\pi\omega\frac{r_+{}^2-r_-{}^2}{r_+-r_-}\big)+\cos\big(2\pi\sqrt{(l+\frac{1}{2})^2-2\omega^2(r_+{}^2+r_+r_-+r_-{}^2)}\big)}{\cosh \big(2\pi\omega\frac{r_+{}^2+r_-{}^2}{r_+-r_-}\big)+\cos\big(2\pi\sqrt{(l+\frac{1}{2})^2-2\omega^2(r_+{}^2+r_+r_-+r_-{}^2)}\big)}.
\label{eq:lowfreq_r}
\end{align}

One can check that $\lvert \widehat{T}\rvert^2+\lvert \widehat{R}\rvert^2=1$ as expected. In addition, if we take the limit $\omega\rightarrow 0$ in the above expressions for  $\lvert \widehat{T}\rvert^2$ and $\lvert \widehat{R}\rvert^2$, we find
\begin{align}
    \lvert \widehat{T}\rvert&\rightarrow\frac{2r_+r_-}{r_+{}^2+r_-{}^2}=\frac{Q^2}{2M^2-Q^2}\\
    \lvert \widehat{R}\rvert&\rightarrow\frac{r_+{}^2-r_-{}^2}{r_+{}^2+r_-{}^2}=\frac{2M\sqrt{M^2-Q^2}}{2M^2-Q^2}.
\end{align}
Interestingly, it is not the case that $\lvert \widehat{T}\rvert \rightarrow 0$ as $\omega\rightarrow 0$, as might naively be expected for the exterior region. This result is consistent with that of \citep{kehle2019}, where the corresponding scattering coefficients in the interior are uniformly bounded.

We will now apply~\eqref{eq:lowfreq_t} and~\eqref{eq:lowfreq_r} to the discussion at the end of Section~\ref{sec:extended_solution}. By comparing~\eqref{eq:lowfreq_t} and~\eqref{eq:lowfreq_r} to~\eqref{eq:scattering_trace_t} and~\eqref{eq:scattering_trace_r}, we see that in our low-frequency approximation,
\begin{align}
    \tfrac{1}{2}\operatorname{tr}M_{++}=-\cos\big(2\pi\sqrt{(l+\tfrac{1}{2})^2-2\omega^2(r_+{}^2+r_+r_-+r_-{}^2)}\big).
    \label{eq:low_frequency_trace}
\end{align}
First, we note that if the cosine equals $\pm 1$, the inequalities in~\eqref{eq:conv_cond} are saturated. For $\omega=0$ we have $\tfrac{1}{2}\operatorname{tr}M_{++}=1$, which is the lower limit of~\eqref{eq:conv_cond}. Thus, for small, nonzero $\omega$ we have $\tfrac{1}{2}\operatorname{tr}M_{++}<1$ and the inequalities~\eqref{eq:conv_cond} are satisfied. Thus, we conclude that the solution space of a maximal analytic extension of RN is non-divergent in the low-frequency limit. In the next section, we use a numerical series solution method to determine whether a non-divergent solution space also exists at mid- to high-frequencies.

\subsection{Numerical Series Solution}

In this section, we use the Frobenius method outlined in Bender and Orszag~\cite{bender1999} to obtain numerical results for the scattering coefficients $\widehat{T}$ and $\widehat{R}$. We expand the radial function $R(r)$ about the singular points $r=r_+$ and $r=r_-$ with the same boundary conditions as the radial mode $R_\text{II}$ in~\eqref{eq:II_basis_1}. By matching the solutions at $r=M$, which is always between the two horizons, we can find numerical results for $|\widehat{T}|^2$ and $|\widehat{R}|^2$.

We begin by proposing an ansatz, which is an expansion about the regular singular point $r=r_+$ or $r=r_-$:
\begin{align}
R(r-r_\pm)=(r-r_\pm)^\alpha\sum_{n=0}^\infty a_n(r-r_\pm)^n,
\label{eq:frobenius_ansatz}
\end{align}
where $\alpha$ is an unspecified constant called the indicial exponent, and the $a_n$ are constant coefficients. We then plug~\eqref{eq:frobenius_ansatz} into~\eqref{eq:radial_orig}, and collect terms in powers of $r-r_\pm$. The coefficient of each power of $r-r_\pm$ must be equal to zero in order to satisfy the resulting equation. By setting the coefficient of \mbox{$(r-r_\pm)^{\alpha-2}$} equal to zero, we find two solutions for the indicial exponent:
\begin{align}
\alpha&=+ \rmi\frac{\omega r_\pm{}^2}{r_+-r_-},
\label{eq:indicial_plus}\\
\alpha &= -\rmi\frac{\omega  r_\pm{}^2}{r_+-r_-}.
\label{eq:indicial_minus}
\end{align}
Using the indicial exponent in~\eqref{eq:indicial_plus}, we can see from Eq~\eqref{eq:radial_interest} that $(r-r_-)^\alpha$ is proportional to $e^{-i\omega r_*}$ when expanding about $r=r_-$, and $(r-r_+)^\alpha$ is proportional to $e^{+i\omega r_*}$ when expanding about $r=r_+$. We then set $a_0=1$ and set the remaining coefficients of $(r-r_\pm)^{\alpha+n-2}$ equal to zero, which results in a five-term recurrence relation for the remaining $a_n$. When numerically evaluating the series in~\eqref{eq:frobenius_ansatz}, we find that the terms exponentially converge to the exact solution. Thus, we can safely numerically evaluate the derivative:
\begin{align}
R'(r-r_\pm)=(r-r_\pm)^\alpha\sum_{n=0}^\infty  a_n n (r-r_\pm)^{n-1}.
\label{eq:frobenius_derivative}
\end{align}
Furthermore, the radius of convergence is at least as large as the distance to the next singular point of~\eqref{eq:radial_orig}, so expansions about $r=r_-$ and $r=r_+$ will both converge at $r=M$. We use this fact to match expansions about $r=r_-$ to the expansion about $r=r_+$ which corresponds to $R_\text{II}$ in~\ref{eq:II_basis_1}.

\begin{figure}
\centering
\includegraphics[trim=40 0 0 0,clip,scale=.5]{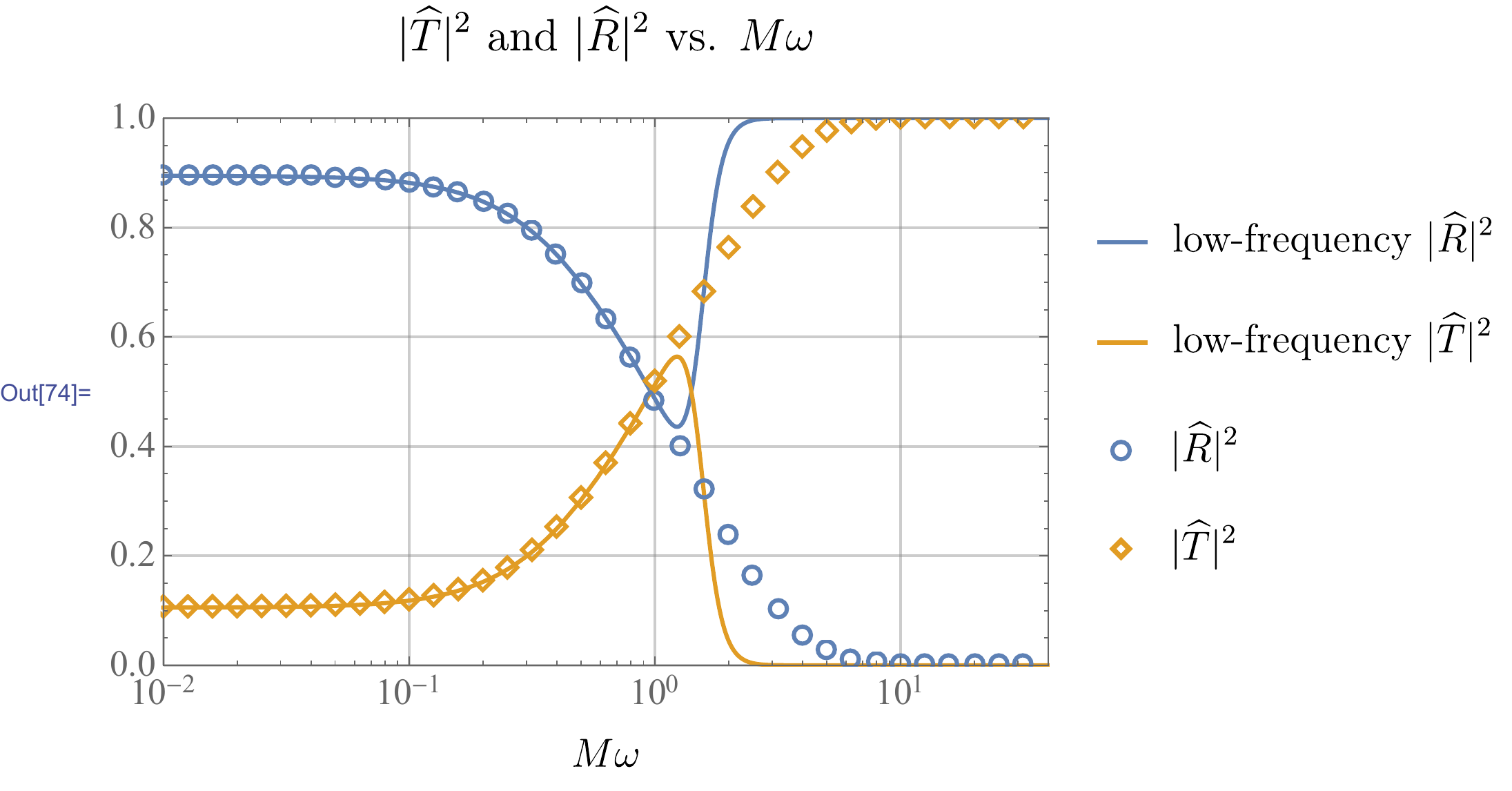}
\caption{A plot of $|\widehat{R}|^2$ and $|\widehat{T}|^2$ vs $M\omega$, obtained by evaluating a Frobenius series for each value of $M\omega$ for fixed $Q/M=0.7$ and $l=2$. The solid lines are the low-frequency approximations for $|\widehat{R}|^2$ and $|\widehat{T}|^2$, given in~\eqref{eq:lowfreq_t} and~\eqref{eq:lowfreq_r}. As expected, the low-frequency approximations break down after $M\omega\approx 1$.}
\label{fig:numerical_rt}
\end{figure}

\begin{figure}
\centering
\includegraphics[trim=40 0 0 0,clip,scale=.5]{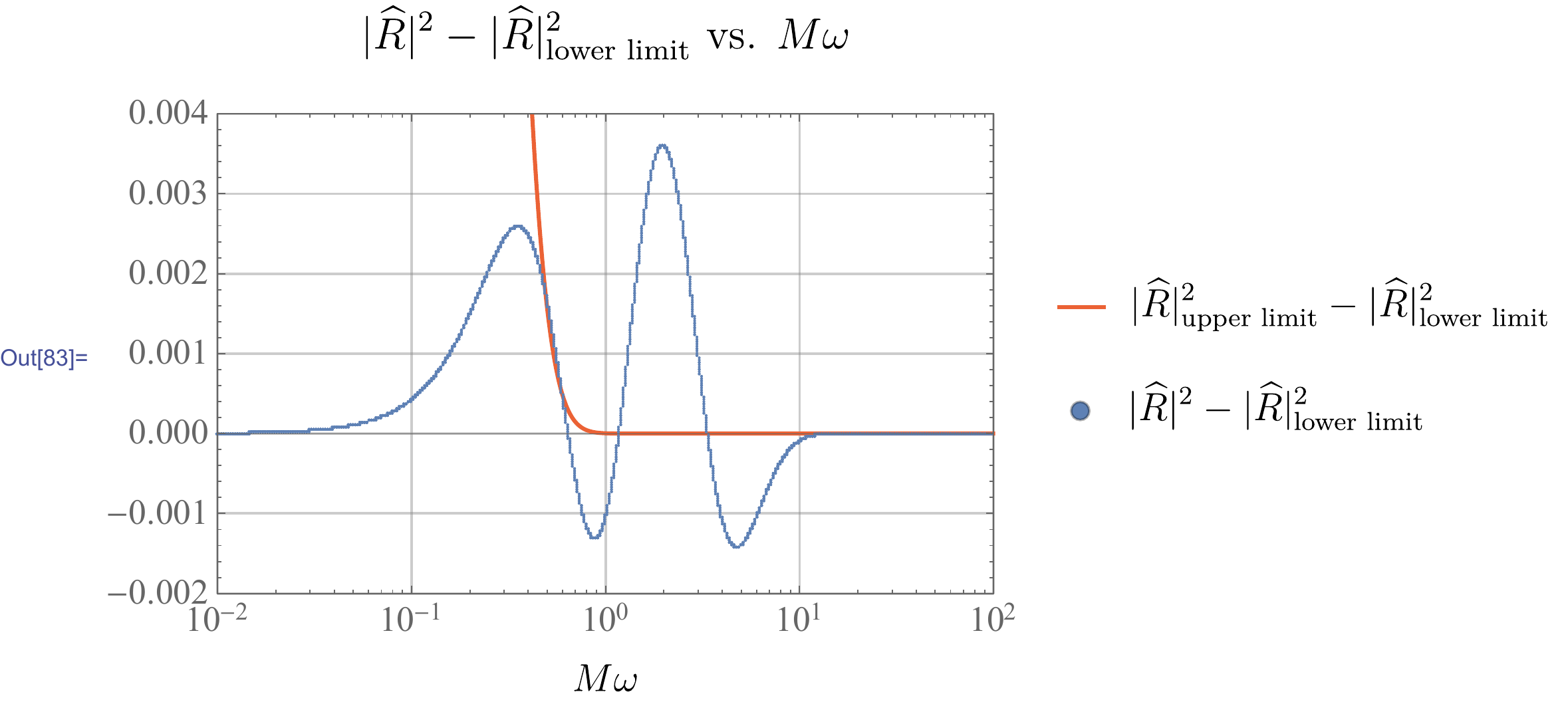}
\caption{The difference between our numerical series result for $|\widehat{R}|^2$ and the lower limit for obtaining complex eigenvalues, shown in~\eqref{eq:conv_cond}. Note that the true values of $|\widehat{R}|^2$ remain near the lower limit for all $M\omega$, not just at low frequency as one might expect. Above the upper limit curve, the upper limit of~\eqref{eq:conv_cond}, and below the horizontal axis, the eigenvalues of $M_{++}$ given in~\eqref{eq:matrix_eigenvalues} are real, resulting in a divergent solution space.}
\label{fig:numerical_difference}
\end{figure}

\begin{figure}
\centering
\includegraphics[trim=40 0 0 0,clip,scale=.5]{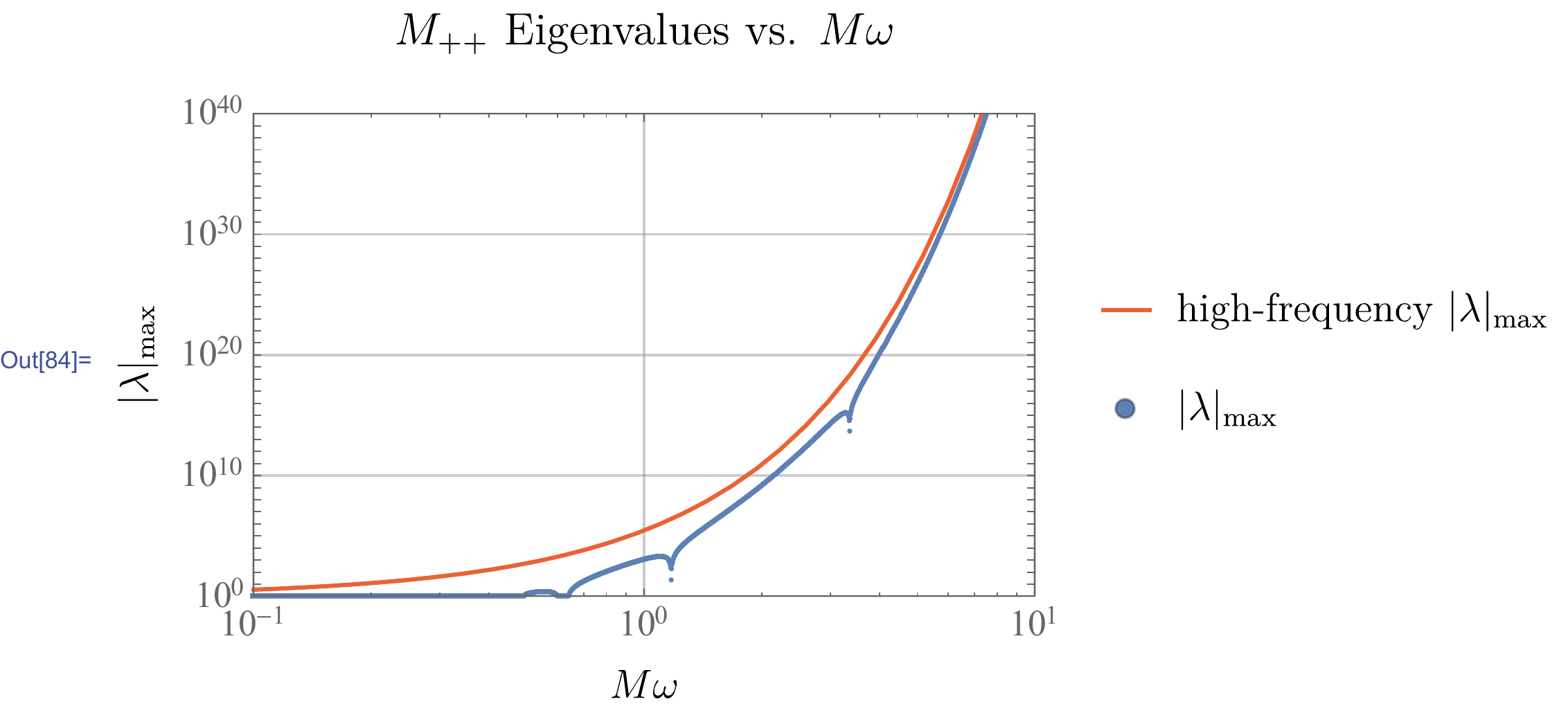}
\caption{A plot of the eigenvalues of $M_{++}$ vs $M\omega$. We can see that when $|\widehat{R}|^2$ lies between the lower and upper limits shown in figure~\ref{fig:numerical_difference}, $|\lambda|=1$, and the solution space is non-divergent. As $M\omega$ increases, $\lvert T \rvert^2\rightarrow 1$, and the eigenvalues approach their high-frequency limit of \mbox{$\lambda_\pm \rightarrow (K_+/K_-)^{\pm 1}=\rme^{\pm 4\pi M\omega}$}. From this plot, we see that the solution space is non-divergent only for certain regions of $M\omega$.}
\label{fig:numerical_eigenvalues}
\end{figure}

We evaluated enough terms in~\eqref{eq:frobenius_ansatz} and~\eqref{eq:frobenius_derivative} in order to obtain 16 digits of accurate data. After matching the expansions at $r/M=1$ for a range of values of $M\omega$ at fixed $Q/M=0.7$ and $l=2$, we obtained the data shown in figure~\ref{fig:numerical_rt}. The low-frequency approximation breaks down around $M\omega\approx 1$ as might be expected. However, the high-frequency behavior of the approximation is determined by the coefficient of the $\omega^2$ term under the square root in~\eqref{eq:lowfreq_r}. While dropping different terms in the differential equation could improve the behavior of our approximation at high frequency, we cannot justify any such changes on the basis of the approximation we have used. For example, when we neglect the $\omega^2$ term, the approximation of $|\widehat{R}|^2$ is equal to the lower limit of~\eqref{eq:conv_cond} and is rather good even at high frequencies, as seen in figure~\ref{fig:numerical_difference}. The figure shows the difference between our numerical result and the lower limit of~\eqref{eq:conv_cond}, which we can use to determine when the eigenvalues are real and when they are complex conjugate pairs of unit modulus. When $|\widehat{R}|^2$ lies between the horizontal axis and the upper limit curve, the eigenvalues are complex. Otherwise, the eigenvalues are real. This can be seen in figure~\ref{fig:numerical_eigenvalues}, where we plot the modulus of the eigenvalues. When the eigenvalues are complex, they have unit modulus. When the eigenvalues are real, they grow exponentially, approaching the high frequency limit, where $|\widehat{T}|^2\rightarrow 1$ and $\lambda_\pm \rightarrow (K_+/K_-)^{\pm 1}=\rme^{\pm 4\pi M\omega}$.

We are now in a position to answer the question posed at the end of section~\ref{sec:extended_solution}: is it possible to construct a non-divergent, globally positive frequency solution for a maximal analytic extension of RN spacetime? The answer is yes, but only for certain ranges of $M\omega$ for fixed $Q/M$ and $l$. In our case study of $Q/M=0.7$ and $l=2$, we can see in figure~\ref{fig:numerical_eigenvalues} that the only values of $M\omega$ which give non-divergent solutions are approximately $M\omega<0.5$ and then again briefly at $M\omega\approx 0.6$, $1.1$, and $3.1$. These values correspond to the regions in figure~\ref{fig:numerical_difference} where $|\widehat{R}|^2$ lies between the horizontal axis and the upper limit. When $\omega$ takes on these values, the global solution acquires only a phase when propagating across each extended geometry, and is thus non-divergent.

\section{Linearly Independent and Antipodal Symmetric Fields}
\label{sec:linear_independence_symmetry}

In the analytically extended domains given in figure~\ref{fig:inner_outer_coordinates}, the antipodal transformation of the coordinates is given by~\citep{sanchez1987}
\begin{align}
    (U_{\pm},V_{\pm},\theta,\phi)\rightarrow(-U_\pm,-V_\pm,\pi-\theta,\phi+\pi)
    \label{eq:antipodal_identification}
\end{align}
The goal of this section is to build solutions which are symmetric or anti-symmetric with respect to this transformation out of the positive and negative frequency solutions. To find these antipodal symmetric and anti-symmetric fields, we first establish how many solutions are linearly independent for a geometry with $n$ horizons, then construct a basis of positive and negative frequency solutions, and finally use this basis to construct the antipodal symmetric solutions.

\subsection{Linear Independence of Positive and Negative Frequency Solutions}

There are two distinctions between global solutions in extended RN which contribute to the number of linearly independent solutions. The first distinction is in the specified boundary conditions, and the second is in the choice of analytic continuation across each bifurcation point (choosing a positive- or negative-frequency solution at that horizon). In this section we sketch an induction argument, which shows that a basis of Klein-Gordon solutions in extended RN has $2(n+1)$ linearly independent solutions, where $n$ is the number of inner and outer horizons, counted together. Equivalently, $n$ is the number of bifurcation points in the Penrose diagram. At the core of this result, it should be noted that extra solutions are
being added at each bifurcation point because initial data in region II$'$
(or II) is not sufficient to describe any solution uniquely in regions I
and I$'$ ( or III and III$'$).  However, as will be seen below, our choice to
continue solutions analytically as positive (or negative) frequency at
each bifurcate point does directly address this inherent freedom.

For the base case $n=0$, we consider region II with no horizons. Because the Klein-Gordon equation is second-order, there are two linearly independent solutions in this region, and no analytic continuation is needed. We've proven the base case.

\begin{figure}
    \centering
    \includegraphics[scale=1]{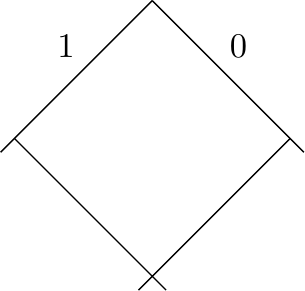}
    \hspace{.5cm}
    \includegraphics{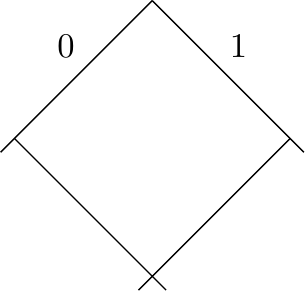}
    \caption{A representation of $2(n+1)$ linearly independent solutions. There are $n+1$ solutions of each form.}
    \label{fig:proofv2_1}
\end{figure}

For the induction step, we imagine that we know there are $2(n+1)$ linearly independent solutions for a solution of $n$ horizons linked together as shown in figure~\ref{fig:reissner_chain}. Here we consider the case where the extended RN geometry terminates with region II, but the argument for a geometry terminating with region II$'$ is nearly identical. Without loss of generality, we can take the solutions to be of the forms in figure~\ref{fig:proofv2_1}. There are $n+1$ solutions of the form on the left and $n+1$ solutions of the form on the right. This specifies the boundary conditions.

\begin{figure}
    \centering
    \begin{minipage}{.23\textwidth}
    \includegraphics[scale=.8]{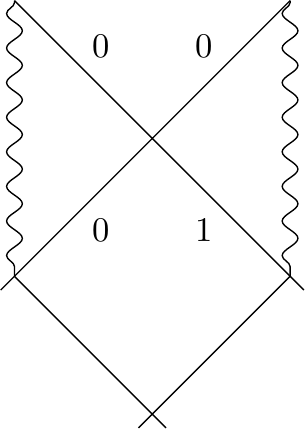}
    
    $\dfrac{K_-{}^2\varphi_{U_-,-}-\varphi_{U_-,+}}{K_-{}^2-1}$
    
    \end{minipage}
    \begin{minipage}{.23\textwidth}
    \includegraphics[scale=.8]{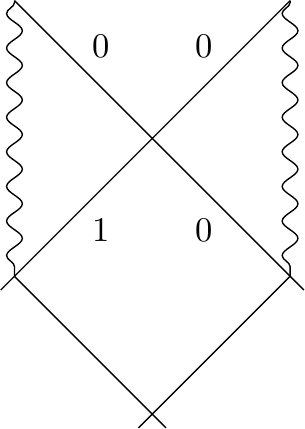}
    
    $\dfrac{K_-{}^2\varphi_{V_-,+}-\varphi_{V_-,-}}{K_-{}^2-1}$

    \end{minipage}
    \begin{minipage}{.23\textwidth}
    \includegraphics[scale=.8]{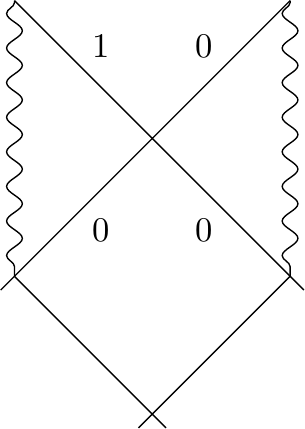}

    $\dfrac{\varphi_{U_-,+}-\varphi_{U_-,-}}{K_--1/K_-}$
    \end{minipage}
    \begin{minipage}{.23\textwidth}
    \includegraphics[scale=.8]{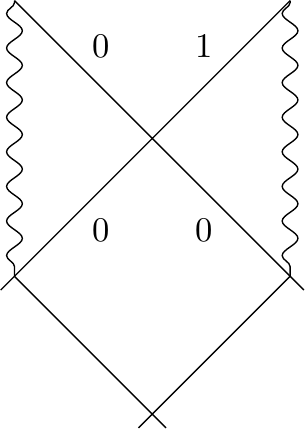}
    
    $\dfrac{\varphi_{V_-,-}-\varphi_{V_+,-}}{K_--1/K_-}$
    \end{minipage}
    \caption{A set of solutions obtained through analytic extension of those represented in figure~\ref{fig:proofv2_1}, using the notation of figure~\ref{fig:horizon_frequency_basis_inner} in~\ref{sec:pos_neg_freq}. The left two diagrams represent the original  $2(n+1)$ solutions, identical to the solutions in figure~\ref{fig:proofv2_1}, except the scattering coefficients on the new uppermost boundaries are equal to zero. The right two diagrams each represent the two ``new'' solutions, each obtained from a linear combination of analytic extensions of the old solutions.}
    \label{fig:proofv2_3}
\end{figure}

Now consider an analytical extension of the geometry with an additional inner horizon. It will suffice to construct a basis of solutions for the new geometry with $2(n+2)$ independent solutions. We extend each solution with positive frequency and with negative frequency and take the linear combinations shown in figure~\ref{fig:proofv2_3}. (See~\ref{sec:pos_neg_freq} for details on the analytic extension procedure.) The first two diagrams represent the original $2(n+1)$ solutions with the scattering coefficients on the new uppermost boundaries equal to zero. The second two diagrams have nonzero coefficients only on the new uppermost boundaries and represent the two ``new'' solutions obtained from the analytical extension. Because the first two diagrams are a basis for all lower boundaries (from the induction hypothesis), and the second two diagrams from a complete basis of the two new independent boundaries, this set of $2(n+2)$ solutions form a complete basis for the entire extended geometry. We've proved the induction step and that the Klein-Gordon equation has $2(n+1)$ linearly independent solutions for an extended RN geometry with $n$ horizons.

\subsection{A complete basis of solutions}
\begin{figure}
\centering
\includegraphics[scale=1]{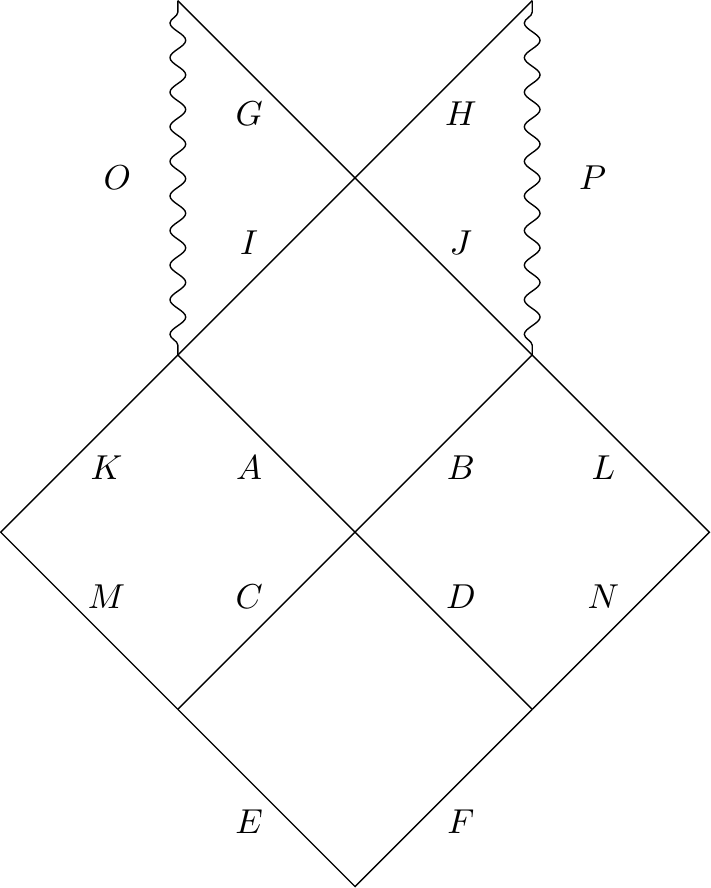}
\caption{An extended RN geometry with all boundaries labeled.}
\label{fig:basis_key}
\end{figure}

In this section (see Table~\ref{tab:basis} for notation), we explicitly construct a basis of solutions for an extended RN geometry with $n=2$ horizons (as shown in figure~\ref{fig:complete_reissner}), which means that there are six linearly independent solutions. The chart in Table~\ref{tab:basis} shows the scattering amplitudes of the solution on the corresponding boundary in figure~\ref{fig:basis_key}.

One may notice that the solutions $\phi^-_{+,\alpha}$ and $\phi_+^-{}_{,\beta}$ are missing from this basis. This is because they are not linearly independent from the other six. To see this, notice that only six of the scattering coefficients for the boundaries in figure~\ref{fig:basis_key} are not determined by scattering. Two of these six coefficients are set boundary conditions, and the other four are determined by the choice of analytic continuation at the respective horizon. We choose boundaries $A$, $B$, $C$, $D$, $G$, and $H$ to be those with the six independent scattering coefficients. Since we have six linearly independent solutions, we can fully determine these six coefficients, including constructing the solutions $\phi^-_{+,\alpha}$ and $\phi^-_{+,\beta}$.

We can write any solution $\phi$ for the extended RN geometry in the form
\begin{align}
\phi&=a\phi^+_{+,\alpha} + b\phi^+_{-,\alpha} + c\phi^-_{-,\alpha}
 + d\phi^+_{+,\beta} + e\phi^+_{-,\beta} + f\phi^-_{-,\beta},
\end{align}
so we can find the scattering coefficients at each boundary from a matrix $M$:
\begin{align}
\begin{pmatrix}
A\\B\\C\\D\\G\\H
\end{pmatrix}
&=
M\begin{pmatrix}
a\\b\\c\\d\\e\\f
\end{pmatrix}\equiv \begin{pmatrix}
0&0&0&1&1&1\\
1&1&1&0&0&0\\
1/K_+&K_+&K_+&0&0&0\\
0&0&0&K_+&1/K_+&1/K_+\\
K_-\widehat{R}/\widehat{T}&K_-\widehat{R}/\widehat{T}&\widehat{R}/K_-\widehat{T}&K_-/\widehat{T}^*&K_-/\widehat{T}^*& 1/K_-\widehat{T}^*\\
1/K_-\widehat{T}&1/K_-\widehat{T}&K_-/\widehat{T}&\widehat{R}^*/K_-\widehat{T}^*&\widehat{R}^*/K_-\widehat{T}^*&K_-\widehat{R}^*/\widehat{T}^*
\end{pmatrix}
\begin{pmatrix}
a\\b\\c\\d\\e\\f
\end{pmatrix},
\label{eq:find_comb}\\
\nonumber \\
M^{-1}&=
\begin{pmatrix}
0& \frac{K_+}{K_+-1/K_+}&-\frac{1}{K_+-1/K_+}&0&0&0\\
\frac{(K_-+1/K_-)\widehat{R}^*}{(K_--1/K_-)\widehat{T}^*{}^2}&g&\frac{1}{K_+-1/K_+}&0&-\frac{\widehat{R}^*}{(K_--1/K_-)\widehat{T}^*}&-\frac{1}{(K_--1/K_-)\widehat{T}^*}\\
-\frac{(K_-+1/K_-)\widehat{R}^*}{(K_--1/K_-)\widehat{T}^*{}^2}&-\frac{1/K_-+K_-\lvert \widehat{R}\rvert^2}{(K_--1/K_-)\lvert \widehat{T}\rvert^2}&0&0&\frac{\widehat{R}^*}{(K_--1/K_-)\widehat{T}^*}&\frac{1}{(K_--1/K_-)\widehat{T}^*}\\
-\frac{1}{K_+(K_+-1/K_+)}&0&0&\frac{1}{K_+-1/K_+}&0&0\\
-g&-\frac{(K_-+1/K_-)\widehat{R}}{(K_--1/K_-)\widehat{T}^2}&0&-\frac{1}{K_+-1/K_+}&\frac{1}{(K_--1/K_-)\widehat{T}}&\frac{\widehat{R}}{(K_--1/K_-)\widehat{T}}\\
\frac{K_-+\lvert \widehat{R}\rvert^2/K_-}{(K_--1/K_-)\lvert \widehat{T}\rvert^2}&\frac{(K_-+1/K_-)\widehat{R}}{(K_--1/K_-)\widehat{T}^2}&0&0& -\frac{1}{(K_--1/K_-)\widehat{T}}&-\frac{\widehat{R}}{(K_--1/K_-)\widehat{T}}
\end{pmatrix},
\end{align}
where
\begin{align}
\det M=-(K_+-1/K_+)^2(K_--1/K_-)^2,
\end{align}
and
\begin{align}
g=\frac{(K_+/K_--K_-/K_+)+(K_+K_--1/K_+K_-)\lvert \widehat{R}\rvert^2}{(K_+-1/K_+)(K_--1/K_-)\lvert \widehat{T}\rvert^2}.
\end{align}
From this, we find how to express $\phi^-_{+,\alpha}$ and $\phi^-_{+,\beta}$ in terms of this basis. For example, since we know the values of the scattering coefficients of the solution $\phi^-_{+,\alpha}$, we can use~\eqref{eq:find_comb} to find the appropriate linear combination:
\begin{align*}
\begin{pmatrix}
a\\b\\c\\d\\e\\f
\end{pmatrix}=
M^{-1}\begin{pmatrix}
0\\1\\1/K_+\\0\\ \widehat{R}/K_-\widehat{T}\\ K_-/\widehat{T}
\end{pmatrix}=
\begin{pmatrix}
1\\-1\\1\\0\\0\\0
\end{pmatrix}.
\end{align*}
Doing this again for $\phi^-_{+,\beta}$, we find
\begin{align}
\phi^-_{+,\alpha}&=\phi^+_{+,\alpha}-\phi^+_{-,\alpha}+\phi^-_{-,\alpha}\\
\phi^-_{+,\beta}&=\phi^+_{+,\beta}-\phi^+_{-,\beta}+\phi^-_{-,\beta}.
\end{align}

\subsection{Constructing a basis of symmetric and antisymmetric solutions}

Using the method outlined in the previous section, we can construct a basis of six solutions, which are symmetric or antisymmetric with respect to the transformation~\eqref{eq:antipodal_identification} at each horizon:
\begin{align}
\begin{pmatrix}
\phi^s_{s,\alpha}\\
\phi^s_{a,\alpha}\\
\phi^a_{a,\alpha}\\
\phi^s_{s,\beta}\\
\phi^s_{a,\beta}\\
\phi^a_{a,\beta}\\
\end{pmatrix}=
S
\begin{pmatrix}
\phi^+_{+,\alpha}\\ \phi^+_{-,\alpha}\\ \phi^-_{-,\alpha}\\
\phi^+_{+,\beta}\\ \phi^+_{-,\beta}\\ \phi^-_{-,\beta}
\end{pmatrix},
\end{align}
where the $a$ or $s$ indicates antisymmetric or symmetric on the given horizon, respectively, and the matrix $S$ is 
\begin{align}
S=\begin{pmatrix}
\frac{K_+}{K_++1}&-\frac{K_+-K_-+\lvert \widehat{R}\rvert^2(1-K_+K_-)}{(K_++1)(K_-+1)\lvert \widehat{T}\rvert^2}& \hphantom{+}\frac{1-K_-\lvert \widehat{R}\rvert^2}{(K_-+1)\lvert \widehat{T}\rvert^2}&0&-\frac{(K_--1)\widehat{R}}{(K_-+1)\widehat{T}^2}&\frac{(K_--1)\widehat{R}}{(K_-+1)\widehat{T}^2}\\
\frac{K_+}{K_+-1}&-\frac{K_++K_--\lvert \widehat{R}\rvert^2(1+K_+K_-)}{(K_+-1)(K_-+1)\lvert \widehat{T}\rvert^2}& \hphantom{+} \frac{1-K_-\lvert \widehat{R}\rvert^2}{(K_-+1)\lvert \widehat{T}\rvert^2}&0&-\frac{(K_--1)\widehat{R}}{(K_-+1)\widehat{T}^2}&\frac{(K_--1)\widehat{R}}{(K_-+1)\widehat{T}^2}\\
\frac{K_+}{K_+-1}&\hphantom{+}\frac{K_+-K_--\lvert \widehat{R}\rvert^2(1-K_+K_-)}{(K_+-1)(K_--1)\lvert \widehat{T}\rvert^2}&- \frac{1+K_-\lvert \widehat{R}\rvert^2}{(K_--1)\lvert \widehat{T}\rvert^2}&0&-\frac{(K_-+1)\widehat{R}}{(K_--1)\widehat{T}^2}&\frac{(K_-+1)\widehat{R}}{(K_-+1)\widehat{T}^2}\\
0&\frac{(K_--1)\widehat{R}^*}{(K_-+1)\widehat{T}^*{}^2}&-\frac{(K_--1)\widehat{R}^*}{(K_-+1)\widehat{T}^*{}^2}&\hphantom{+}\frac{1}{K_++1}&\frac{K_+-K_-+\lvert \widehat{R}\rvert (1-K_+K_-)}{(K_++1)(K_-+1)\lvert \widehat{T}\rvert^2}&\frac{K_--\lvert \widehat{R}\rvert^2}{(K_-+1)\lvert \widehat{T}\rvert^2}\\
0&\frac{(K_--1)\widehat{R}^*}{(K_-+1)\widehat{T}^*{}^2}&-\frac{(K_--1)\widehat{R}^*}{(K_-+1)\widehat{T}^*{}^2}&-\frac{1}{K_+-1}&\frac{K_++K_--\lvert \widehat{R}\rvert (1+K_+K_-)}{(K_+-1)(K_-+1)\lvert \widehat{T}\rvert^2}&\frac{K_--\lvert \widehat{R}\rvert^2}{(K_-+1)\lvert \widehat{T}\rvert^2}\\
0&\frac{(K_-+1)\widehat{R}^*}{(K_--1)\widehat{T}^*{}^2}&-\frac{(K_-+1)\widehat{R}^*}{(K_--1)\widehat{T}^*{}^2}&-\frac{1}{K_+-1}&-\frac{K_+-K_--\lvert \widehat{R}\rvert (1-K_+K_-)}{(K_+-1)(K_--1)\lvert \widehat{T}\rvert^2}&\frac{K_-+\lvert \widehat{R}\rvert^2}{(K_--1)\lvert \widehat{T}\rvert^2}
\end{pmatrix}.
\end{align}
The symmetric and antisymmetric solutions are shown explicitly in Table~\ref{tab:basis_symmetric}. The remaining two symmetric/antisymmetric solutions can be written in terms of these six:
\begin{align}
\phi^a_{s,\alpha}&=\phi^s_{s,\alpha}-\phi^s_{a,\alpha}+\phi^a_{a,\alpha}\\
\phi^a_{s,\beta}&=\phi^s_{s,\beta}-\phi^s_{a,\beta}+\phi^a_{a,\beta}.
\end{align}


Note that even though the scattering coefficients to the past and to the future match at the two inner horizons, one cannot simply ``stack'' these solutions in order to obtain symmetric solutions for an extended geometry. Instead, one must construct $2n+2$ solutions with pure frequency at each of the $n$ horizons, and then invert the \mbox{$(2n+2)\times(2n+2)$} matrix representing those solutions in order to obtain the symmetric solutions. This is clearly no easy task, but is in principle possible for any finite number of horizons.

\begin{table}
\small
\caption{Six solutions that form a complete basis for an extended RN geometry. The upper sign indicates the frequency on the inner horizon, and the lower sign indicates the frequency on the upper horizon. The boundary conditions are applied in region II, and the two distinct conditions are signified by $\alpha$ and $\beta$.  The letters $A$ through $P$ in the first column corresponds to the boundary in figure~\ref{fig:basis_key}, and the table entries give the scattering coefficient of the corresponding boundary and solution.}
\vspace{.5cm}

\begin{tabular}{lllllll}
\br
  & $\phi^+_{+,\alpha}$ & $\phi^+_{-,\alpha}$ & $\phi^-_{-,\alpha}$& $\phi^+_{+,\beta}$ & $\phi^+_{-,\beta}$ & $\phi^-_{-,\beta}$ \\ \mr
  
$A$ & $0$ & $0$ & $0$ & $1$ & $1$ & $1$\\
$B$ & $1$ & $1$ & $1$ & $0$ & $0$ & $0$\\ 
$C$ & $1/K_+$ & $K_+$ & $K_+$ & $0$ & $0$ & $0$ \\ 
$D$ & $0$ & $0$ & $0$ & $K_+$ & $1/K_+$ & $1/K_+$ \\ 
$E$ & $\widehat{R}/K_+\widehat{T}$ & $K_+\widehat{R}/\widehat{T}$ & $K_+\widehat{R}/\widehat{T}$ & $K_+/\widehat{T}^*$ & $1/K_+\widehat{T}^*$ & $1/K_+\widehat{T}^*$\\ 
$F$ & $1/K_+\widehat{T}$ & $K_+/\widehat{T}$ & $K_+/\widehat{T}$& $K_+\widehat{R}^*/\widehat{T}^*$ & $\widehat{R}^*/K_+\widehat{T}^*$ & $\widehat{R}^*/K_+\widehat{T}^*$ \\ 
$G$ & $K_-\widehat{R}/\widehat{T}$ & $K_-\widehat{R}/\widehat{T}$ & $\widehat{R}/K_-\widehat{T}$ & $K_-/\widehat{T}^*$ & $K_-/\widehat{T}^*$ & $1/K_-\widehat{T}^*$\\ 
$H$ & $1/K_-\widehat{T}$ & $1/K_-\widehat{T}$ & $K_-/\widehat{T}$& $\widehat{R}^*/K_-\widehat{T}^*$ & $\widehat{R}^*/K_-\widehat{T}^*$ & $K_-\widehat{R}^*/\widehat{T}^*$\\ 
$I$ & $1/\widehat{T}$ & $1/\widehat{T}$ & $1/\widehat{T}$ & $\widehat{R}^*/\widehat{T}^*$ & $\widehat{R}^*/\widehat{T}^*$ & $\widehat{R}^*/\widehat{T}^*$ \\ 
$J$ & $\widehat{R}/\widehat{T}$ & $\widehat{R}/\widehat{T}$ & $\widehat{R}/\widehat{T}$  & $1/\widehat{T}^*$ & $1/\widehat{T}^*$ & $1/\widehat{T}^*$ \\ 
$K$ & $1/K_+T$ & $K_+/T$ & $K_+/T$ & $R^*/T^*$ & $R^*/T^*$ & $R^*/T^*$  \\ 
$L$ & $R/T$ & $R/T$ & $R/T$& $K_+/T^*$ & $1/K_+T^*$ & $1/K_+T^*$ \\ 
$M$ & $R/K_+T$ & $K_+R/T$ & $K_+R/T$& $1/T^*$ & $1/T^*$ & $1/T^*$  \\ 
$N$ & $1/T$ & $1/T$ & $1/T$& $K_+R^*/T^*$ & $R^*/K_+T^*$ & $R^*/K_+T^*$  \\ 
$O$ &  \pbox{20cm}{$\frac{K_-\widehat{R}B^*-B}{i\widehat{T}/2\omega r_+}$ \\ $-\frac{K_-\widehat{R}A^*-A}{i\widehat{T}/2\omega r_+}\frac{r}{r_-}$} &  \pbox{20cm}{$\frac{K_-\widehat{R}B^*-B}{i\widehat{T}/2\omega r_+}$ \\ $-\frac{K_-\widehat{R}A^*-A}{i\widehat{T}/2\omega r_+}\frac{r}{r_-}$} & \pbox{20cm}{$\frac{\widehat{R}B^*/K_--B}{i\widehat{T}/2\omega r_+}$ \\ $-\frac{\widehat{R}A^*/K_--A}{i\widehat{T}/2\omega r_+}\frac{r}{r_-}$}  &\pbox{20cm}{$\frac{K_-B^*-\widehat{R}^*B}{i\widehat{T}^*/2\omega r_+}$ \\ $-\frac{K_-A^*-\widehat{R}^*A}{i\widehat{T}^*/2\omega r_+}\frac{r}{r_-}$} &   \pbox{20cm}{$\frac{K_-B^*-\widehat{R}^*B}{i\widehat{T}^*/2\omega r_+}$ \\ $-\frac{K_-A^*-\widehat{R}^*A}{i\widehat{T}^*/2\omega r_+}\frac{r}{r_-}$} &  \pbox{20cm}{$\frac{B^*/K_--\widehat{R}^*B}{i\widehat{T}^*/2\omega r_+}$ \\ $-\frac{A^*/K_--\widehat{R}^*A}{i\widehat{T}^*/2\omega r_+}\frac{r}{r_-}$} \\ 
$P$ &  \pbox{20cm}{$\frac{K_-\widehat{R}B^*-B}{i\widehat{T}K_-/2\omega r_+}$ \\ $-\frac{K_-\widehat{R}A^*-A}{i\widehat{T}K_-/2\omega r_+}\frac{r}{r_-}$} &  \pbox{20cm}{$\frac{K_-\widehat{R}B^*-B}{i\widehat{T}K_-/2\omega r_+}$ \\ $-\frac{K_-\widehat{R}A^*-A}{i\widehat{T}K_-/2\omega r_+}\frac{r}{r_-}$} & \pbox{20cm}{$\frac{\widehat{R}B^*-BK_-}{i\widehat{T}/2\omega r_+}$ \\ $-\frac{\widehat{R}A^*-K_-A}{i\widehat{T}/2\omega r_+}\frac{r}{r_-}$} & \pbox{20cm}{$\frac{K_-B^*-\widehat{R}^*B}{i\widehat{T}^*K_-/2\omega r_+}$ \\ $-\frac{K_-A^*-\widehat{R}^*A}{i\widehat{T}^*K_-/2\omega r_+}\frac{r}{r_-}$} & \pbox{20cm}{$\frac{K_-B^*-\widehat{R}^*B}{i\widehat{T}^*K_-/2\omega r_+}$ \\ $-\frac{K_-A^*-\widehat{R}^*A}{i\widehat{T}^*K_-/2\omega r_+}\frac{r}{r_-}$} &  \pbox{20cm}{$\frac{B^*-K_-\widehat{R}^*B}{i\widehat{T}^*/2\omega r_+}$ \\ $-\frac{A^*-K_-\widehat{R}^*A}{i\widehat{T}^*/2\omega r_+}\frac{r}{r_-}$}\\ \br
\end{tabular}
\label{tab:basis}
\end{table} 


\newcommand{\specialcell}[2][c]{
  \begin{tabular}[#1]{@{}c@{}}#2\end{tabular}}
  
\begin{table}
\small
\caption{Six solutions, which are antisymmetric (signified by $a$) or symmetric (signified by $s$) at each horizon. The upper letter is the type of symmetry on the inner horizon, and the lower letter is the type of symmetry on the outer horizon. The $\alpha$ or $\beta$ signifies the boundary condition chosen in region II, as in the solutions shown in Table~\ref{tab:basis}. The letters $A$ through $P$ in the first column correspond to the boundary in figure~\ref{fig:basis_key}, and the table entries give the scattering coefficient of the corresponding boundary and solution.}
\vspace{.5cm}
\begin{tabular}{lllllll}
\br
 & $\phi^s_{s,\alpha}$ & $\phi^s_{a,\alpha}$ & $\phi^a_{a,\alpha}$ & $\phi^s_{s,\beta}$ & $\phi^s_{a,\beta}$ & $\phi^a_{a,\beta}$ \\
\mr
$A$ & $0$ & $0$ & $0$ & $1$ & $1$ & $1$\\ 
$B$ & $1$ & $1$ & $1$ & $0$& $0$& $0$ \\ 
$C$ & $1$ & $-1$ & $-1$ & $0$& $0$& $0$  \\ 
$D$ & 0 & 0 & 0 & $1$& $-1$& $-1$\\ 
$E$ & $\widehat{R}/\widehat{T}$ & $-\widehat{R}/\widehat{T}$ & $-\widehat{R}/\widehat{T}$ & $1/\widehat{T}^*$& $-1/\widehat{T}^*$& $-1/\widehat{T}^*$ \\
$F$ & $1/\widehat{T}$ & $-1/\widehat{T}$ & $-1/\widehat{T}$ & $\widehat{R}^*/\widehat{T}^*$& $-\widehat{R}^*/\widehat{T}^*$& $-\widehat{R}^*/\widehat{T}^*$\\ 
$G$ & $\widehat{R}/\widehat{T}$ & $\widehat{R}/\widehat{T}$ & $-\widehat{R}/\widehat{T}$ & $1/\widehat{T}^*$& $1/\widehat{T}^*$& $-1/\widehat{T}^*$\\
$H$ & $1/\widehat{T}$ & $1/\widehat{T}$ & $-1/\widehat{T}$ & $\widehat{R}^*/\widehat{T}^*$ & $\widehat{R}^*/\widehat{T}^*$ & $-\widehat{R}^*/\widehat{T}^*$\\
$I$ & $1/\widehat{T}$ & $1/\widehat{T}$ & $1/\widehat{T}$ & $\widehat{R}^*/\widehat{T}^*$ & $\widehat{R}^*/\widehat{T}^*$ & $\widehat{R}^*/\widehat{T}^*$\\
$J$ & $\widehat{R}/\widehat{T}$ & $\widehat{R}/\widehat{T}$ & $\widehat{R}/\widehat{T}$& $1/\widehat{T}^*$& $1/\widehat{T}^*$& $1/\widehat{T}^*$ \\ 
$K$ & $1/T$ & $-1/T$ & $-1/T$ & $R^*/T^*$& $R^*/T^*$& $R^*/T^*$ \\ 
$L$ & $R/T$ & $R/T$ & $R/T$ & $1/T^*$& $-1/T^*$& $-1/T^*$\\ 
$M$ & $R/T$ & $-R/T$ & $-R/T$ & $1/T^*$& $1/T^*$& $1/T^*$\\ 
$N$ & $1/T$& $1/T$& $1/T$ & $R^*/T^*$& $-R^*/T^*$& $-R^*/T^*$\\ 
$O$ & \pbox{20cm}{$\frac{B^*\widehat{R}-B}{i\widehat{T}/2r_+\omega}$\\ $-\frac{A^*\widehat{R}-A}{i\widehat{T}/2r_+\omega}\frac{r}{r_-}$} & \pbox{20cm}{$\frac{B^*\widehat{R}-B}{i\widehat{T}/2r_+\omega}$\\ $-\frac{A^*\widehat{R}-A}{i\widehat{T}/2r_+\omega}\frac{r}{r_-}$} & \pbox{20cm}{$-\frac{B^*\widehat{R}+B}{i\widehat{T}/2r_+\omega}$\\ $+\frac{A^*\widehat{R}+A}{i\widehat{T}/2r_+\omega}\frac{r}{r_-}$}& \pbox{20cm}{$\frac{B^*-B\widehat{R}^*}{i\widehat{T}^*/2r_+\omega}$ \\ $-\frac{A^*-A\widehat{R}^*}{i\widehat{T}^*/2r_+\omega}\frac{r}{r_-}$}&\pbox{20cm}{$\frac{B^*-B\widehat{R}^*}{i\widehat{T}^*/2r_+\omega}$ \\ $-\frac{A^*-A\widehat{R}^*}{i\widehat{T}^*/2r_+\omega}\frac{r}{r_-}$}& \pbox{20cm}{$-\frac{B^*+B\widehat{R}^*}{i\widehat{T}^*/2r_+\omega}$ \\ $+\frac{A^*+A\widehat{R}^*}{i\widehat{T}^*/2r_+\omega}\frac{r}{r_-}$}\\
$P$ &  \pbox{20cm}{$\frac{B^*\widehat{R}-B}{i\widehat{T}/2r_+\omega}$\\ $-\frac{A^*\widehat{R}-A}{i\widehat{T}/2r_+\omega}\frac{r}{r_-}$} & \pbox{20cm}{$\frac{B^*\widehat{R}-B}{i\widehat{T}/2r_+\omega}$\\ $-\frac{A^*\widehat{R}-A}{i\widehat{T}/2r_+\omega}\frac{r}{r_-}$} & \pbox{20cm}{$\frac{B^*\widehat{R}+B}{i\widehat{T}/2r_+\omega}$\\ $-\frac{A^*\widehat{R}+A}{i\widehat{T}/2r_+\omega}\frac{r}{r_-}$} & \pbox{20cm}{$\frac{B^*-B\widehat{R}^*}{i\widehat{T}^*/2r_+\omega}$ \\ $-\frac{A^*-A\widehat{R}^*}{i\widehat{T}^*/2r_+\omega}$}& \pbox{20cm}{$\frac{B^*-B\widehat{R}^*}{i\widehat{T}^*/2r_+\omega}$ \\ $-\frac{A^*-A\widehat{R}^*}{i\widehat{T}^*/2r_+\omega}\frac{r}{r_-}$}& \pbox{20cm}{$\frac{B^*+B\widehat{R}^*}{i\widehat{T}^*/2r_+\omega}$ \\ $-\frac{A^*+A\widehat{R}^*}{i\widehat{T}^*/2r_+\omega}\frac{r}{r_-}$}\\ \br 
\end{tabular}
\label{tab:basis_symmetric}
\end{table}

\section{Discussion}
\label{sec:discussion}

In this paper we have developed the classical tools necessary to construct an antipodal symmetric QFT in RN in analogy to~\citep{schroedinger2011,sanchez1987_2}. See also~\citep{domenech1987,stephens1989,friedman1995,chang2007} for related constructions in Rindler, de Sitter and anti-de Sitter spacetimes. Moreover, we have uncovered rich features of the Klein-Gordon equation in RN along the way. We have characterized Klein-Gordon solutions in terms of their scattering data (transmission and reflection amplitudes) in each of regions I, II, and III respectively, and used this data to construct globally positive (and negative) frequency solutions throughout the entire domain portrayed in figure~\ref{fig:complete_reissner}.  By careful choice of an initial starting point, we could characterize these solutions in terms of the surface gravity at each of the two horizons, and the scattering data arising just in region II. 
Motivated by this fact and an infinite analytic extension of the type shown in figure~\ref{fig:reissner_chain}, we also sought the low frequency limits of, and numerical results for, $|\widehat{T}|$ and $|\widehat{R}|$ in order to determine the field amplitude as it propagates along the analytic extension. We found that, for low enough frequency, field amplitudes of solutions with purely positive or negative frequency at each horizon will acquire only a phase after passing both the inner and outer horizons, while at higher frequencies the amplitudes will tend to grow exponentially in one direction, and decay in the other direction.

The positive and negative frequency we have constructed lend themselves to a thermal description because their analyticity reflects periodicity in imaginary time with a temperature at each horizon fixed by its surface gravity.  This characteristic is precisely that recognized by Israel \citep{israel1976} to bring out the correspondence between black hole thermodynamics and the independently developed notions of thermo-field theory.  This initially arose for completely different reasons but, in this context, hints at the possibility of antipodal identification, may permit a microcanonical representation of entropy for eternal black holes \cite{Martinez:1994ub}, and leads to the possibility of generalizing the path integral approach to black-hole thermodynamics \cite{Frolov:1994gz}.

Ideas of antipodal identification were originally proposed by 't Hooft in relation to the Schwarzschild black hole~\citep{tHooft:1983kru,tHooft:1984kcu}, and led to the specific construction of globally symmetric or antisymmetric fields under the antipodal map by others~\citep{sanchez1987}. Antipodal identification is still favoured for some formulations related to quantum considerations. For quantum fields, in particular, the globally symmetric or antisymmetric constructions are best expressed in terms of the globally positive and negative frequency solutions, which can be associated with individual particle states. Here we have extended the related constructions to the Reissner-Nordstr\"om  spacetime, which necessitates a somewhat richer analysis because the analytic extensions of the spacetime may contain an arbitrarily large number of horizons and spacetime regions.

Our construction has been especially oriented toward its potential future application to quantum fields in the maximally extended RN spacetime.  As we added additional asymptotically null boundaries (or additional horizons), we necessarily increased the diversity of field configurations which needed to be considered. For an extended geometry with $n$ horizons, we have demonstrated how to construct a basis consisting of $2(n+1)$ linearly independent symmetric and antisymmetric fields. We have done this explicitly for $n=2$. In order to construct this basis, we need solutions with positive frequency on some horizons and negative frequency on others. Once identified, this symmetric/antisymmetric basis could be used to begin constructing the corresponding QFT.

\section*{Acknowledgements}  
ATF and BFW wish to express gratitude to Gerard 't Hooft for initially inspiring this project.
ATF acknowledges partial support for this work by FCT/Portugal through UID/MAT/04459/2013, grant (GPSEinstein) PTDC/MAT-ANA/1275/2014 and SFRH/BPD/115959/2016. BFW acknowledges support from NSF grants PHY-1205906, PHY-1314529 and PHY-1607323. NAS acknowledges support from the Graduate Student Fellowship from the University of Florida.
\vfill\eject

\bibliographystyle{iopart-num}
\bibliography{references}

\pagebreak

\appendix

\section{Positive and Negative Frequency Solutions}
\label{sec:pos_neg_freq}

\begin{figure*}[t]

\begin{minipage}{.48\textwidth}
\centering
\includegraphics[scale=.8]{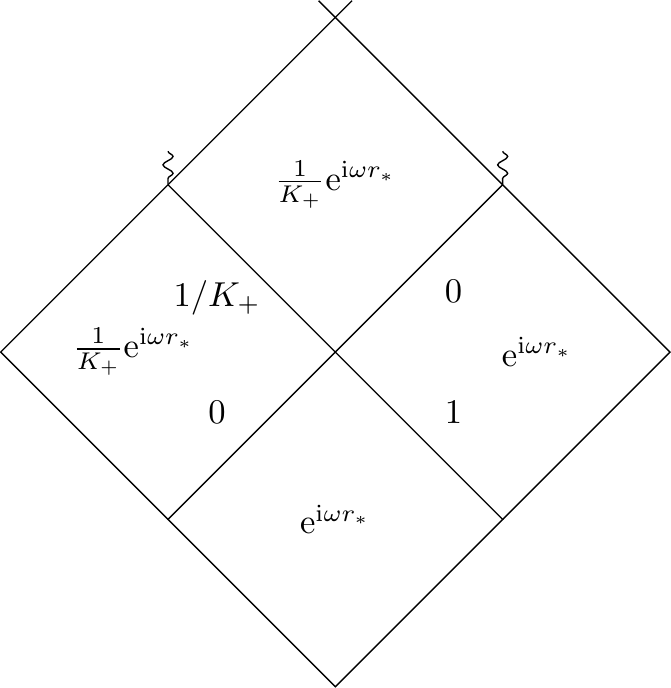}

$\varphi_{U_+,+}$
\end{minipage}
\begin{minipage}{.48\textwidth}
\centering
\includegraphics[scale=.8]{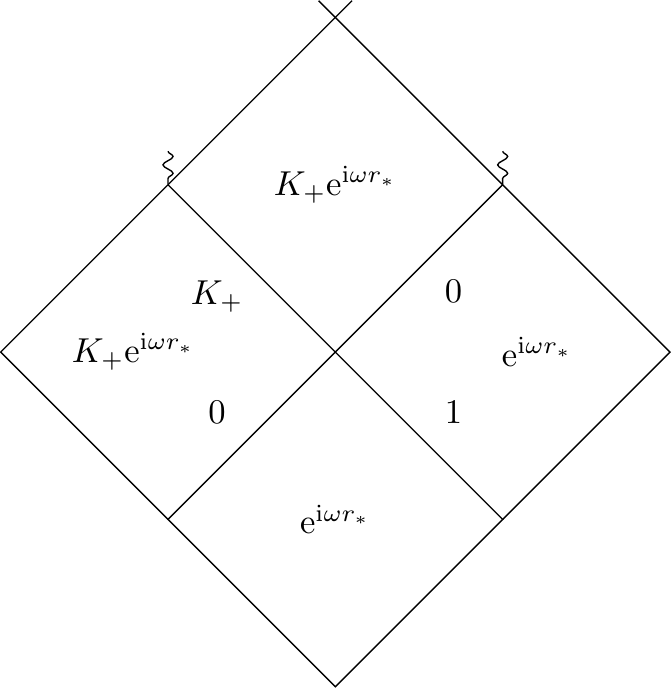}

$\varphi_{U_+,-}$
\end{minipage}
\begin{minipage}{.48\textwidth}
\vspace{.5cm}
\centering
\includegraphics[scale=.8]{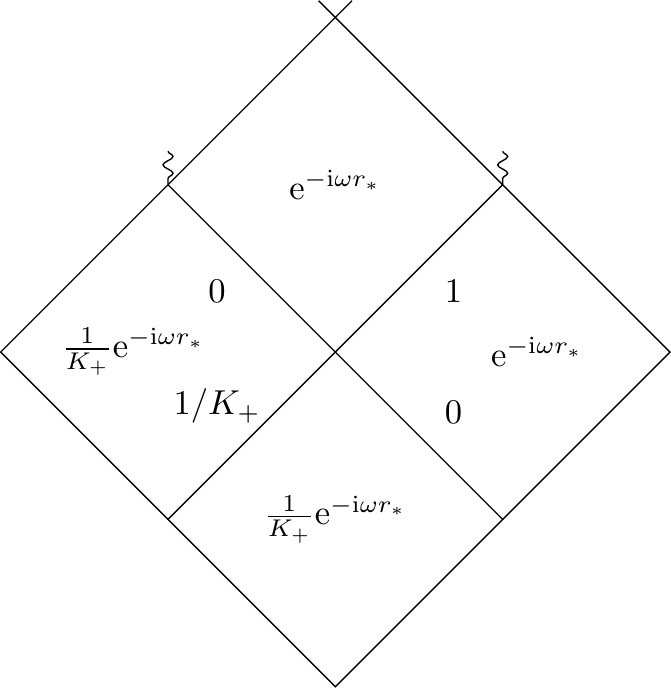}

$\varphi_{V_+,+}$
\end{minipage}\hspace{.4cm}
\begin{minipage}{.48\textwidth}
\vspace{.5cm}
\centering
\includegraphics[scale=.8]{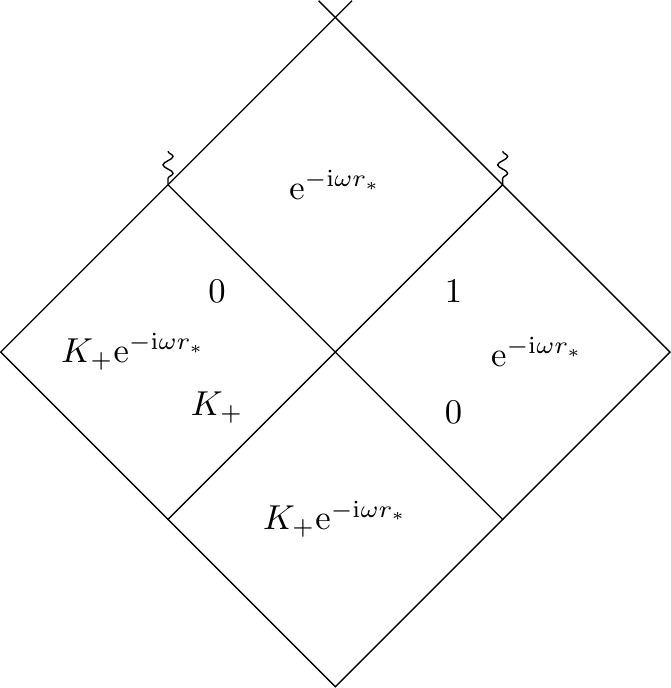}

$\varphi_{V_+,-}$
\end{minipage}
\caption{Four pure-frequency Klein-Gordon solutions for the outer horizon, the $U_+$ or $V_+$ signifying the boundary conditions and the subscript $\pm$ indicating positive- or negative-frequency. The scattering amplitudes are labeled on each horizon, and the radial part of the solution is written in the center of each region.}
\label{fig:horizon_frequency_basis_outer}
\end{figure*}

In developing positive- and negative-frequency solutions to the Klein-Gordon equation, we restrict our attention to the behavior of solutions near the horizons, starting with the outer horizon $r=r_+$. Using the development in section~\ref{sec:klein_gordon}, by specifying boundary conditions in region I of figure~\ref{fig:complete_reissner}, we can construct two linearly independent solutions \mbox{$\varphi_{U_+}\approx \frac{r_+}{r}e^{i\omega(r_*-t)}Y_{lm}$} and $\varphi_{V_+}\approx \frac{r_+}{r} e^{-i\omega(r_*+t)}Y_{lm}$, which are defined near the past and future outer horizons, respectively. We can analytically extend these solutions to regions II, I$'$, and II$'$ by expressing them in terms of the coordinates $U_+$ and $V_+$ given in~\eqref{eq:outer_coordinates_u} and~\eqref{eq:outer_coordinates_v}:
\begin{align}
    \varphi_{U_+}&\approx \frac{r_+}{r}U_+{}^{\rmi\omega/\kappa_+}Y_{lm},\\
    \varphi_{V_+}&\approx \frac{r_+}{r} V_+{}^{-\rmi\omega/\kappa_+}Y_{lm}.
\end{align}
One can think of $\varphi_U$ as waves crossing the past horizon and $\varphi_V$ as waves crossing the future horizon. We would like to express $\varphi_{U_+}$ and $\varphi_{V_+}$ in terms of $r_*$ and $t$ in regions II, I$'$ and II$
'$, which will allow us to construct global solutions. This is accomplished by specifying $-1=e^{\pm i\pi}$ in~\eqref{eq:outer_coordinates_u} and~\eqref{eq:outer_coordinates_v}:
\begin{align}
    \varphi_{U_+}|_{\text{I},\text{II}'}&=\frac{r_+}{r}|U_+|^{\rmi\omega/\kappa_+}Y_{lm} \nonumber \\
        &= \frac{r_+}{r}\rme^{\rmi\omega(r_*-t)}Y_{lm}\\
    \varphi_{U_+}|_{\text{I}',\text{II}}&=\frac{r_+}{r} (-|U_+|)^{\rmi\omega/\kappa_+}Y_{lm} \nonumber \\
        &=K_+{}^{\pm 1}\frac{r_+}{r}\rme^{\rmi\omega(r_*-t)}Y_{lm}
        \label{eq:uplus_change}\\
    \varphi_{V_+}|_{\text{I,II}}&= \frac{r_+}{r}|V_+|^{-\rmi\omega/\kappa_+}Y_{lm}\nonumber \\
        &= \frac{r_+}{r} \rme^{-\rmi\omega(r_*+t)}Y_{lm}\\
    \varphi_{V_+}|_{\text{I}',\text{II}'}&= \frac{r_+}{r}( -|V_+|)^{-\rmi\omega/\kappa_+}Y_{lm} \nonumber\\
        &=K_+{}^{\mp 1} \frac{r_+}{r} \rme^{-\rmi\omega(r_*+t)}Y_{lm},
        \label{eq:vplus_change}
\end{align}
where $K_+=e^{\pi \omega/\kappa_+}$. For convenience, we absorb the factor of $K_+$ into the radial part of our solutions. See figure~\ref{fig:horizon_frequency_basis_outer} for a graphical representation of these solutions.

It turns out that the four analytic continuations are each purely postive- or negative-frequency. The terminology is borrowed from the flat space case, where positive-frequency solutions are defined as those whose time derivative pulls out a factor of $-i\omega$ and negative-frequency solutions pull out a factor of $+i\omega$. This is not a covariant definition for the curved-space case ~\citep{carroll2004}. A covariant norm definition consistent with the flat space definition uses the following covariant indefinite inner product, which is an integral on a global, spacelike (Cauchy) surface:

\begin{align}
    \langle \phi,\varphi\rangle &= -\rmi\int (\phi^*\partial_\mu\varphi- \varphi\partial_\mu\phi^*) \sqrt{-g}\, \rmd\Sigma^\mu,
    \label{eq:norm}
\end{align}
which is invariant to the choice of surface. Positive-frequency solutions are positive-norm solutions and negative-frequency solutions are negative-norm solutions ~\citep{carroll2004}. The sign of the integral in~\eqref{eq:norm} is of course sensitive to the orientation of the Cauchy surface. We choose an orientation such that the normal vector points in the $\partial(V_+-U_+)$ direction in both regions I and I$'$ (to the future on the Penrose diagram). This results in a positive norm for all of our solutions in region I and a negative norm in region I$'$. The choice of sign in the exponents of ~\eqref{eq:uplus_change} and~\eqref{eq:vplus_change} determine the norm of the global solution. For either solution, we find choosing the minus sign in the exponents will result in a globally positive norm because the solution in region I$'$ is smaller in magnitude than the solution in region I. Thus, the solution has a globally positive norm and is positive-frequency on the given horizon. Likewise, the choice of the plus sign for either solution will result in a globally negative norm and thus negative-frequency solution. Graphical representations of these solutions are in figure~\ref{fig:horizon_frequency_basis_outer}.

\begin{figure*}[t]    \centering
    \begin{minipage}{.23\textwidth}
    \centering
    \includegraphics[scale=.8]{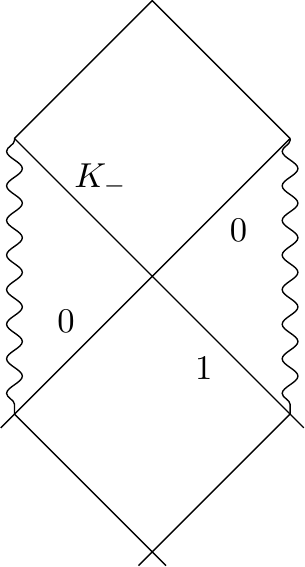}
    
    $\varphi_{U_-,+}$
    \end{minipage}
    \begin{minipage}{.23\textwidth}
    \centering
    \includegraphics[scale=.8]{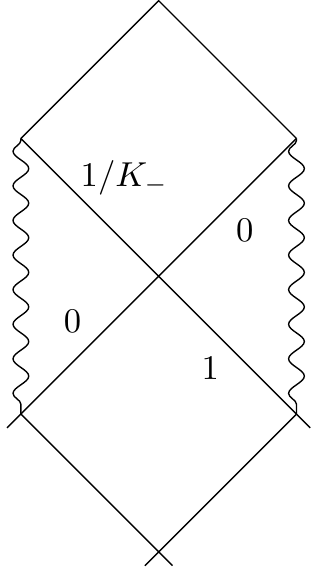}
    
    $\varphi_{U_-,-}$
    \end{minipage}
    \begin{minipage}{.23\textwidth}
    \centering
    \includegraphics[scale=.8]{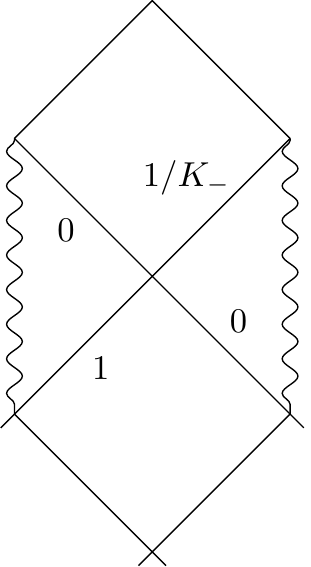}
    
    $\varphi_{V_-,+}$
    \end{minipage}
    \begin{minipage}{.23\textwidth}
    \centering
    \includegraphics[scale=.8]{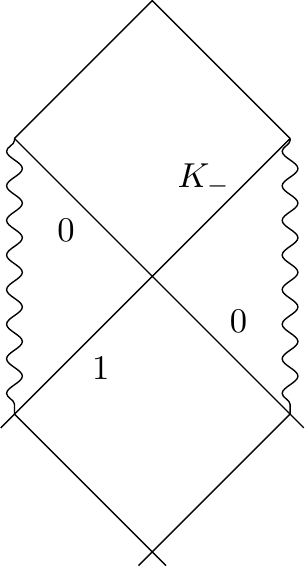}
    
    $\varphi_{V_-,-}$
    \end{minipage}
    \caption{Four pure-frequency Klein-Gordon solutions for the inner horizon, the $U_-$ or $V_-$ signifying the boundary conditions and the subscript $\pm$ indicating positive- or negative-frequency. The scattering amplitudes are labeled on each horizon.}
    \label{fig:horizon_frequency_basis_inner}
\end{figure*}

The development for analytic continuation past the inner horizon is very similar, except we use a normal vector for the Cauchy surface pointing in the $-\partial(V_-+U_-)$ direction in regions III and III$'$ (still to the future on the Penrose diagram). Thus, solutions in region III have positive norm and solutions in region III$'$ have negative norm. We can use this to construct the solutions in figure~\ref{fig:horizon_frequency_basis_inner}, analogous to those in figure~\ref{fig:horizon_frequency_basis_outer}.

We have constructed a global basis of solutions for behavior near the inner and outer horizons of the extended RN manifold. For the past and future horizons each, we have a positive- and negative-frequency analytical continuation to the extended manifold, for a total of four solutions per bifurcation point, represented in figure~\ref{fig:horizon_frequency_basis_outer}. Because we have four solutions and four boundaries at each bifurcation point, we can express a solution with arbitrary scattering amplitudes at these four boundaries in terms of purely positive- and negative-frequency solutions, a fact we use in section~\ref{sec:linear_independence_symmetry}. We use these solutions implicitly in section~\ref{sec:extended_solution} to analytically continue an arbitrary solution past the bifurcation point, choosing from the positive-frequency set. We have now demonstrated how to analytically continue Klein-Gordon solutions past both the inner  and outer horizons with either positive or negative frequency.

\newpage

\end{document}